\documentclass{emulateapj}
\usepackage[utf8]{inputenc}

\usepackage{amsmath}
\usepackage{wasysym}
\usepackage{commath}
\usepackage{bm}
\usepackage{color}

\newcommand{\vo}{\vec{o}\@ifnextchar{^}{\,}{}}

\usepackage[backref,breaklinks,colorlinks,citecolor=blue]{hyperref}        
\usepackage[all]{hypcap}

\slugcomment{Accepted for Publication in ApJ, March 13th, 2017}

\begin{document}
\title{TIME DEPENDENT MODELS OF MAGNETOSPHERIC ACCRETION ONTO YOUNG STARS}

\author{C. E. Robinson\altaffilmark{1}}
\author{J. E. Owen\altaffilmark{2}\altaffilmark{,3}}
\author{C. C. Espaillat\altaffilmark{1}}
\author{F. C. Adams\altaffilmark{4}}

\shortauthors{Robinson et al.}
\shorttitle{MAGNETOSPHERIC ACCRETION ONTO YOUNG STARS}

\email{connorr@bu.edu}
\altaffiltext{1}{Department of Astronomy, Boston University, 725 Commonwealth Avenue, Boston, MA 02215, USA}
\altaffiltext{2}{Institute for Advanced Study, Einstein Drive, Princeton, NJ 08540, USA}
\altaffiltext{3}{Hubble Fellow}
\altaffiltext{4}{Physics Department, University of Michigan, Ann Arbor, MI 48109, USA}

\begin{abstract}
Accretion onto Classical T Tauri stars is thought to take place through the action of magnetospheric processes, with gas in the inner disk being channeled onto the star's surface by the stellar magnetic field lines. Young stars are known to accrete material in a time-variable manner and the source of this variability remains an open problem, particularly on the shortest ($\sim$ day) timescales. 
Using one-dimensional time-dependent numerical simulations that follow the field line geometry, we find that for plausibly realistic young stars, steady-state transonic accretion occurs naturally in the absence of any other source of variability. 
However, we show that if the density in the inner disk varies smoothly in time with $\sim$ day long time-scales (e.g., due to turbulence) this complication can lead to the development of shocks in the accretion column. These shocks propagate along the accretion column and ultimately hit the star, leading to rapid, large amplitude changes in the accretion rate. We argue that when these shocks hit the star the observed time-dependence will be a rapid increase in accretion luminosity followed by a slower decline and could be an explanation for some of the short period variability observed in accreting young stars.
Our one-dimensional approach bridges previous analytic work to more complicated, multi-dimensional simulations, and observations. 

\end{abstract}
\section{INTRODUCTION}
Excess IR emission in classical T Tauri stars (CTTS) is attributed to optically thick circumstellar disks composed of dust and gas that form alongside the star from the collapsing natal molecular cloud \citep[e.g.][]{mendoza66,shu87}. 
Typical full (or primordial) circumstellar dust disks have inner radii at the dust sublimation radius, roughly 0.05-0.5 AU \citep{natta01, muzerolle03}, and extend out to tens or hundreds of AU \citep[e.g.][]{odell94, isella09, andrews09}.
The gaseous portion of the disk is thought to extend further in to where the stellar magnetic fields are strong enough to disrupt the disk \citep[e.g.,][]{koenigl91,livio92,hartmannbook98,muzerolle03}. 
This occurs roughly at the co-rotation radius, where the Keplerian rotational velocity is equal to the stellar rotation rate (typically 5-10 stellar radii).
Although polarization measurements and modeling of T Tauri stars have shown the magnetic field is complicated near the photosphere \citep{valenti04,donati11b,gregory11}, the large-scale magnetic structure of the star is dominated by its lowest order multipole components. 

In addition to excess IR emission, CTTS exhibit UV luminosities above photospheric levels from shocks caused by material accreting onto the star.  \citep[e.g.,][]{gullbring98,muzerolle98, muzerolle01,muzerolle03, ingleby11, ingleby15}. 
In the canonical magnetospheric accretion scenario, disk material travels from the magnetic truncation radius, along the magnetic field lines, and falls at super sonic free-fall speeds on stellar surface \citep{koenigl91, calvet92, shu94, hartmann94}.
Since the material is supersonic, it shocks as it collides with the star, and heats the gas to temperatures of order $\sim10^5-10^6$~K, which in turn leads to UV and soft X-ray emission \citep[][]{calvet98}. The accretion rate onto the star can be measured by characterizing and modeling this excess emission \citep[][]{gullbring98,muzerolle98,ardila13}.

CTTS are known to be variable, hosting stochastic and quasi-periodic changes in luminosity that occur over a vast range of timescales. Measurements exhibiting variability have been taken across the electromagnetic spectrum, e.g., bursts of X-rays and increased UV luminosity as a result of rapid increases in the accretion rate \citep{ingleby15}, periodic decreases in optical flux explained as warps in the disk variably blocking the star and starspots \citep{bouvier99, alencar10}, as well as `seesaw' variability in the infrared due to changes of the inner disk wall height \citep{muzerolle09,espaillat11}.
Furthermore, equivalent widths and shapes of spectral lines have been observed to change \citep[e.g.,][]{johns95}. 
Longer period changes, on the timescales of months and years, have been observed in CTTS and explained as variable star spot coverage and perhaps hints of stellar activity cycles analogous to the solar cycle \citep{cohen04}. 
All of the aforementioned processes can occur simultaneously, making it difficult to isolate the driving force for a given source of variability.


In this paper, we focus on the innermost region of the disk and variability due to changes in the accretion rate that occur on day to week timescales. 
This type of variability is faster than what is expected from viscous processes since the viscous timescale in the inner disk is on the order of several years \citep[][]{hartmannbook98}.
Proposed sources of variability on these timescales include Rayleigh-Taylor instabilities that produce variable ``tongues" of material that flow between magnetic field lines \citep{kulkarni08} and magnetorotational instabilities that can cause changes in the amount of material available for accretion as a function of time \citep{romanova12}. 
A monitoring campaign of NGC 2264 observed multiple accretion signatures that appear to fit these models well \citep{stauffer14}, providing motivation for continued modeling of the inner regions of the disk. 

The hydrodynamics of gas traveling along magnetic accretion columns has been studied analytically, assuming steady-state flows \citep[e.g.,][]{hartmann94, li96,koldoba02}. Under the assumption that the accretion columns are well described by a polytropic equation of state ($P\propto\rho^{1+1/n}$), analytic work suggests a transonic steady state solution that approaches free-fall velocities may not be possible within the accretion columns if the effective polytropic index, $n$ is smaller than a critical value \citep{adams12}. 
These authors developed a coordinate system that follows magnetic field lines aligned with the stellar rotation axis under the assumption that the magnetic field is strong enough to dominate the flow.
This allows for the transformation of the system to a one-dimensional (1D) problem with semi-analytic steady state solutions for the isothermal case where $n \rightarrow \infty$
The general form for the predicted constraint on the minimum value of $n$ that allows these solutions to exist is $n > \ell + 3/2$, where $\ell$ is the highest order multipole component of the magnetic field near the star. 
This constraint only holds in the inner limit where $r \rightarrow 0$.
The predicted absence of this type of solution under certain conditions may imply the possibility of spontaneous variability generated within accretion columns, even in the case of a constant accretion rate feeding the columns. However, this prediction cannot be tested analytically, and we must appeal to time-dependant numerical calculations. 

Expanding upon this analytic work, we present a 1D hydrodynamic simulation of accretion onto a CTTS using a coordinate construction that follows the accretion columns from the gas disk to the star.
We investigate two possible sources of short timescale accretion variability: inherent variability from the inability to reach free-fall speeds suggested by \citet{adams12} and variability driven by turbulence/instabilities at the inner disk edge. In this work, we also study the hydrodynamics within the accretion column and derive possible observable quantities.

This paper is structured as follows. The construction of the magnetospheric accretion system and our numerical method used to simulate the system are presented in \S 2 and \S 3. We then present simulations testing for spontaneous variability and discuss their implications in \S 4. Explicit time dependence is then included, the results of which are shown in \S 5. We discuss implications of our simulation on the current understanding of variability within the inner disk and possibilities for observations in \S 6 and conclude in \S 7.

\section{MODEL CONSTRUCTION}

The magnetic fields of CTTS are strong near the star, in some cases reaching up to $6\mbox{kG}$ at the photosphere \citep{valenti04}, and as such, we assume that the material in the accretion columns follow the stellar magnetic field lines, which are unperturbed by the accretion flow and co-rotate with the star \citep[see][for a discussion of the validity of this approximation]{koldoba02,adams12}. The magnetic truncation radius, and thus the inner edge of the disk, is assumed to be co-located with the co-rotation radius for simplicity. The large scale magnetic field of CTTS is dominated by the dipole component, but some CTTS have significant octupole components \citep[][]{donati11b}. 
In this work, we investigate two field geometries: a pure dipole field and a field formed from the sum of a dipole and an octupole component. Finally, we take the magnetic field to be constant in time and its axis to be aligned with the rotation axis.

This assumption does prevent the magnetic field lines from becoming twisted from differential rotation, which is an important part of angular momentum transfer within the inner disk \citep{ustyugova06} and could perhaps be a source of variability itself. 
In regards to the transfer of angular momentum, the timescale in which the rotation period of the star changes ($\sim 10^6$ years) \citep{ustyugova06} is many orders of magnitude longer than the roughly week long timescales this paper focuses on.  Thus, we assume the stellar rotation rate is constant.

\subsection{Coordinate System}
Under the assumptions outliend above, the system can be reduced to a 1D problem using a set of coordinates that follows current-free, axis-symmetric magnetic field lines that co-rotate with the star. Because the magnetic field configuration is current free, and thus curl-free, it can be described as the gradient of scalar fields, or coordinates, denoted here as $(p,q)$. 
The coordinate $p$ describes the distance along a given field line, while $q$ describes which field line is being traced. These coordinates are orthogonal; $\nabla p$ points along the field line and $\nabla q$ points in the poloidal direction, perpendicular to $\nabla p$.
The origin of the system is located at the center of the star. A complete formulation of this coordinate system and its associated scale factors can be seen in \citet{adams12}.
For our numerical calculations, a grid was constructed along the magnetic field lines (the $p$ coordinate). A magnetic field containing dipole and octupole contributions is given by

\begin{align}
\begin{split}
\mathbf{B} = \frac{B_{\mbox{oct}}}{2}\xi^{-5}[(5\cos^2\theta - 3)\cos\theta\, \bm{\hat{r}} \\ 
+\frac{3}{4}(5\cos^2\theta - 1)\sin\theta\,\bm{\hat{\theta}}]\\
+\frac{B_{\mbox{dip}}}{2}\xi^{-3}(2\cos{\theta}\,\bm{\hat{r}} + \sin\theta\,\bm{\hat{\theta}})
\end{split}
\end{align}

\noindent where $\xi$ the radius from the center of the star given in dimensionless stellar radius units, $\xi \equiv r/R_*$ and $\theta$ the colatitude. 
$B_{dip}$ and $B_{oct}$ are the strengths of the dipole and octupole terms in the multipole expansion of the magnetic field at the stellar surface.
The parameter $\Gamma$ is defined as the ratio of the coefficients of the composite multipole field, 
\begin{equation}
\Gamma \equiv \frac{B_{\mbox{oct}}}{B_{\mbox{dip}}}.
\end{equation}
The coordinates $q$ and $p$ can be written in terms of the standard spherical polar coordinates.
The coordinate $p$, which traces along individual magnetic field lines, can be written as
\begin{equation}
p = -\frac{1}{4}\xi^{-4}\Gamma (5\cos^2\theta-3)\cos\theta - \xi^{-2}\cos\theta\,,
\end{equation}

\noindent while the orthogonal coordinate $q$ is given by
\begin{equation}
q = \frac{1}{4} \xi^{-3}\Gamma(5\cos^{2}\theta -1)\sin^{2}\theta +\xi^{-1}\sin^{2}\theta.
\end{equation}

\begin{figure}[ht]
\centering
\includegraphics[width = .4\paperwidth]{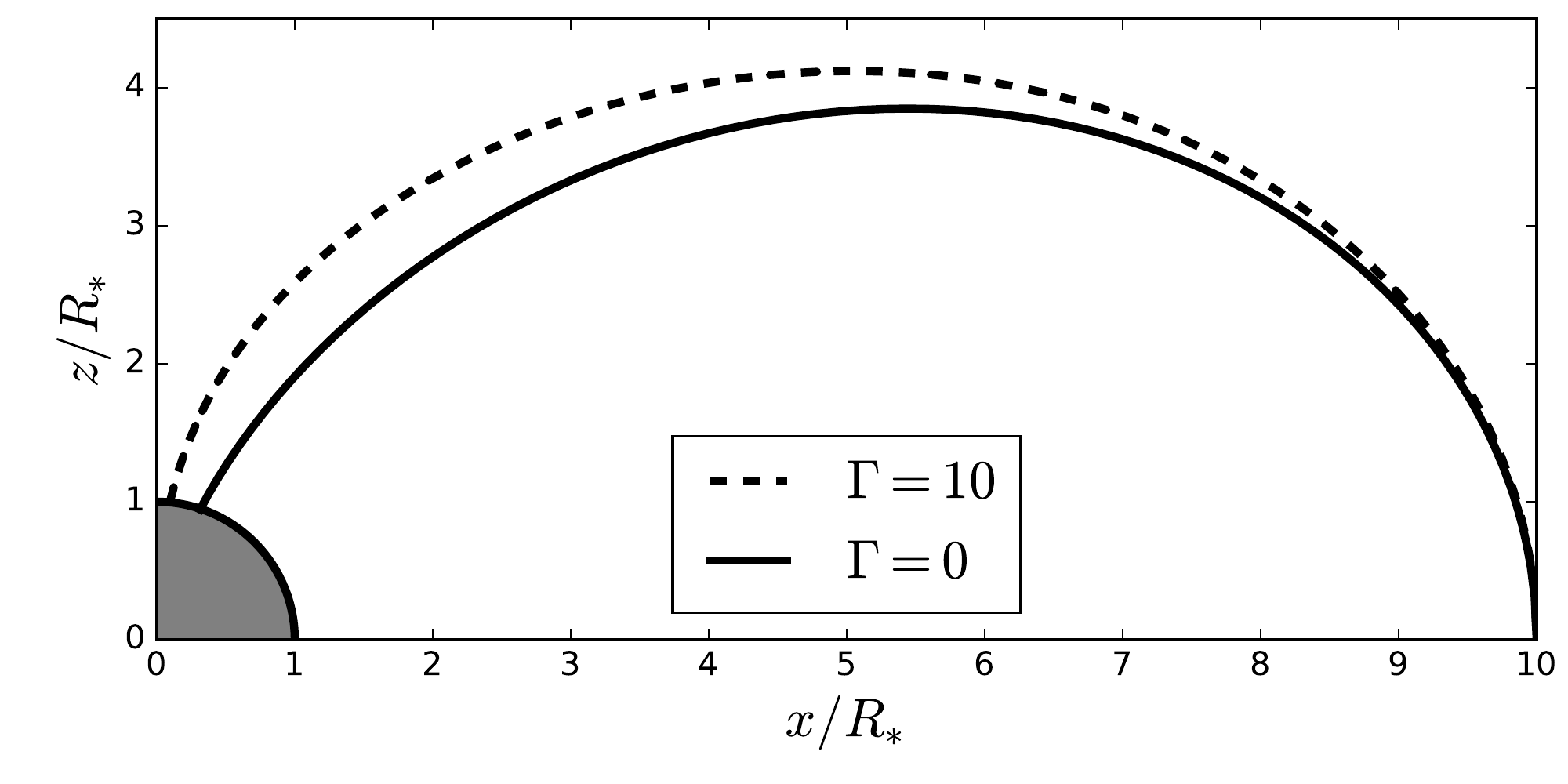}
\includegraphics[width = .4\paperwidth]{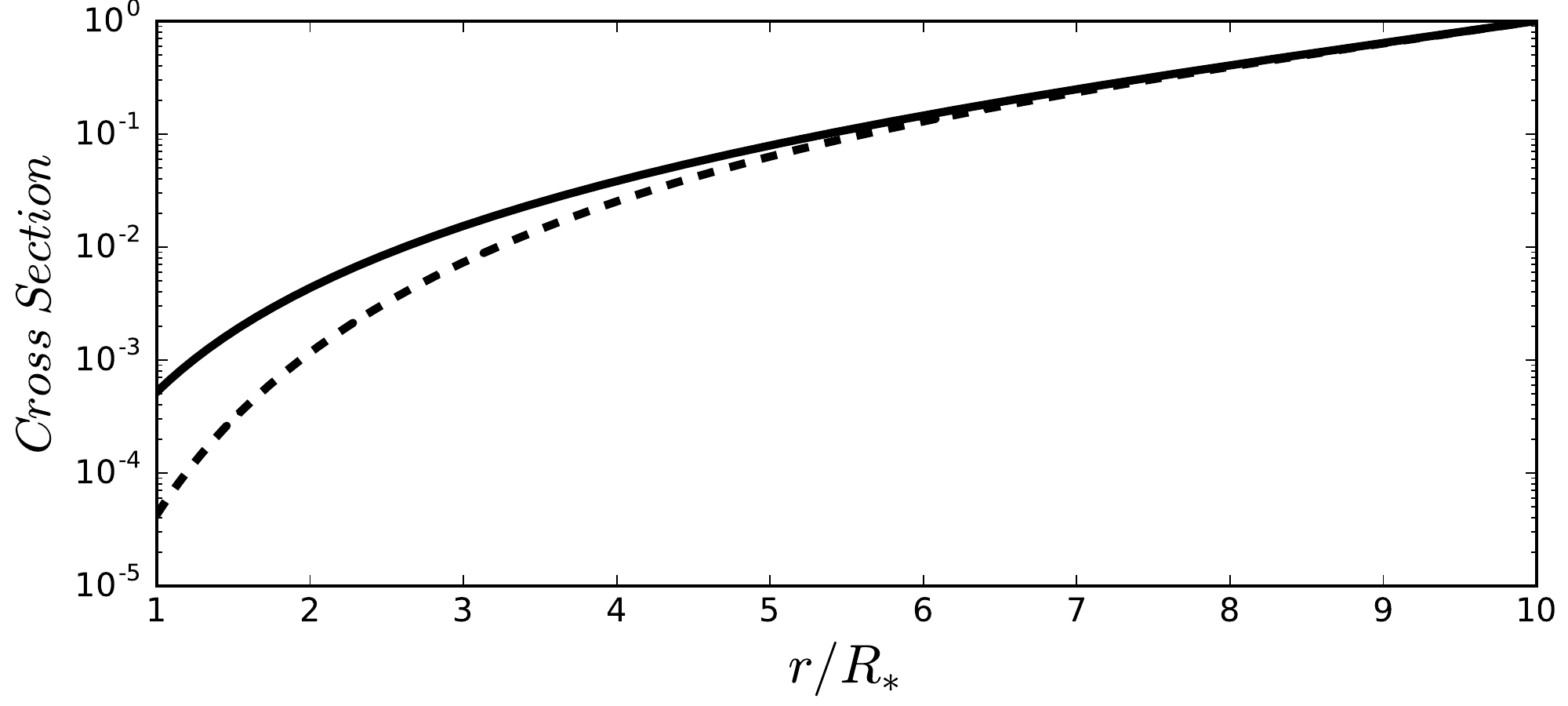}
\caption{\textit{Top panel:} Schematic showing both magnetic field configurations. The solid line traces the field line described by pure dipole fields ($\Gamma = 0$), while the dashed line traces the field line composed of the sum of dipole and octupole components  ($\Gamma = 10$). \textit{Bottom panel:} Cross-sectional area of the accretion column as function of radius, normalized  to the cross-sectional area of the column at the disk. The dipole + octupole field configuration has a smaller cross-section near the stellar surface, leading to higher levels of compression as material flows toward the star compared to the pure dipole field configuration.}
\label{fig:geometry}

\end{figure}

Two configurations of magnetic fields were used in our simulation, $\Gamma = 0$ and $\Gamma = 10$. 
Examples of the field lines for the different geometries are shown in Figure \ref{fig:geometry}. 
The addition of the octupole components has a substantial impact on the regions of the accretion column at small radii. 
Near the surface of the star, the cross-sectional area of the accretion column is greatly reduced in the dipole plus octupole case compared to the pure dipole configuration, which will lead to increased levels of compression of the flow. 
Near the surface of the disk, the effect of adding octupole terms is minimal due to the steep radial fall off of the octupole component, causing the truncation radius to be set almost purely by the strength of the dipole field alone \citep{gregory16}.

\subsection{Hydrodynamic Formulation}
As the magnetic field is assumed to be strong enough to force material to only flow along field lines, the magnetic term in the force equation can be eliminated.
The angular velocity vector is defined such that $\mathbf{\Omega} = \Omega \hat{z}$, and is aligned with the dipole and octupole moments. 
Transforming into a non-inertial frame introduces centrifugal force terms. 
Because the fluid is forced to flow solely along the field lines with no $\hat{\phi}$ components in this frame, the Coriolis term can be eliminated in the 1D force equation. 
After these simplifications, the continuity, force, and energy equations take the form 
\begin{align}
\begin{split}
    \frac{\partial\rho}{\partial t} + \nabla \cdot (\rho \mathbf{u}) = 0\\
    \frac{\partial \mathbf{u}}{\partial t} + (\mathbf{u} \cdot \nabla)\mathbf{u} = -\frac{1}{\rho}\nabla P- \nabla \Psi - \mathbf{\Omega} \times (\mathbf{\Omega} \times \mathbf{r})\\
    \frac{\partial}{\partial t} (\frac{e}{\rho}) + (\mathbf{u} \cdot \nabla)(\frac{e}{\rho}) = -\frac{P}{\rho}\nabla \cdot \mathbf{u}
    \label{eqn:fluideqns}
\end{split}
\end{align}
where $\rho$ is the density, $e$ is the internal energy, $\mathbf{u}$ is the fluid velocity $\Psi$ is the gravitational potential and $P$ is pressure.
Pressure, written in terms of an effective adiabatic index $\gamma = (n+1)/n$ and the ideal gas law, takes the form
\begin{equation}
P = (\gamma-1)e\\
\end{equation}
The sound speed $a$, derived from the adiabatic equation of state and assuming an ideal hydrogen gas, is written as
\begin{equation}
a = \sqrt{\gamma\frac{k_b T}{m_H}}
\end{equation}

\noindent where $k_b$ is the Boltzmann constant, and $m_H$ is the mass of a hydrogen atom, and $T$ is the gas temperature.

The governing equations can be written in dimensionless 1D forms along the co-rotating field line. Reference quantities were used to scale $\rho$, $e$, and the velocity along the field line, $u_p$, written as
\begin{gather}
\alpha \equiv \frac{\rho}{\rho_1}
\quad
u \equiv \frac{u_p}{a_1}
\quad
\epsilon \equiv \frac{e}{a_1^2\rho_1}
\end{gather}

\noindent where $a_1$ and $\rho_1$ are the sound speed and density measured at the inner edge of the disk. Distances have been written in terms of the stellar radius ($\xi$) and time has been scaled using the stellar crossing time for a velocity $a_1$. This allows us to write the governing equations as

\begin{eqnarray}
\frac{\partial \alpha}{\partial t} &=& - \frac{\alpha u}{h_p}\Big[\frac{1}{\alpha}\frac{\partial \alpha}{\partial p} + \frac{1}{u}\frac{\partial u}{\partial p} + \frac{1}{h_q h_\phi}\frac{\partial}{\partial p} (h_q h_\phi)\Big]\\
\frac{\partial u}{\partial t} &=& - \frac{1}{h_p}\Big[u\frac{\partial u}{\partial p} - \frac{1}{\alpha} \frac{\partial P}{\partial p} - \frac{b}{\xi^{7}}\frac{\cos{\theta}f}{H}\Big]\nonumber\\&& +\omega \sin{\theta}(\hat{x}\cdot\hat{p})\\
\frac{\partial}{\partial t} \Big(\frac{\epsilon}{\alpha}\Big) &=& -\frac{u}{h_p}\Big[\frac{\partial}{\partial p}\Big(\frac{\epsilon}{\alpha}\Big) - \frac{(\gamma - 1)}{u}\frac{\epsilon}{\alpha}\frac{\partial u}{\partial p}\Big]\nonumber \\
&& -\frac{u}{h_p}\Big[\frac{(\gamma-1)}{h_q h_\phi}\frac{\epsilon}{\alpha}\frac{\partial}{\partial p}(h_q h_\phi)\Big]
\end{eqnarray}
where $\omega$ is a dimensionless parameter that measures stellar rotation, $\omega = \Omega R_*/a_1$.
The gravitational potential is represented using the dimensionless quantity $b$, which measures the depth of the gravitational potential well relative to the sound speed, written as $b = GM_*/R_*a_{1}^2$. More discussion of this dimensionless formulation of this system is shown in \citep{adams12}.
The curvilinear coordinate scale factor for the $p$ coordinate is denoted $h_p$, while $f$ and $H$ are ancillary functions of $\xi$ and $\theta$. The functional forms for these terms, along with an analytic expression for $(\hat{x}\cdot\hat{p})$ are shown in the appendix.

\section{NUMERICAL METHOD}

The hydrodynamic problem is solved using the method of finite differencing under a {\sc zeus}-style framework using a time-explicit, multi-step solution procedure (see \citealt{stone92}). This style of simulation uses a staggered mesh framework in which the density, internal energy and gravitational potential are zone centered while the velocity is face centered. The grid is constructed with `ghost' cells at both ends of the simulation domain, which are used to set boundary conditions.  
The velocity, density, and internal energy are updated using operator splitting in two sub-steps: the `source' update and the `transport' update.

Our numerical method is based on \citet{owen16}, who simulated magnetically controlled isothermal outflows from hot Jupiters. In this work, the code has been modified to include the energy equation in the system of hydrodynamic equations. In addition, we have modified the geometry of the problem from a dipole plus vertical field to either a pure dipole or a dipole plus octupole magnetic field configuration and added the effects of rotation. More discussion of the changes can be found in the Appendix. 

\subsection{Sub-Steps}
The first step is the source update, in which the velocity and the internal energy are updated by solving finite-difference versions of the governing differential fluid equations. The velocity is updated during the source update due to gravitational forces and pressure gradients along the accretion column.
Because the system is in a non-inertial rotating reference frame, the velocity is also updated due to the centrifugal terms that appear in the force equation. The full difference equations are presented in the Appendix.
In order to capture shocks an artificial viscosity is introduced using a full stress tensor formalism. The full stress tensor is need due to our complicated geometry.
The full details of this artificial viscosity tensor are shown in Appendix A of \citet{owen16}.
Finally, the internal energy of the system is updated due to compressional heating in a time-centered manner \citep[][]{stone92}.

The second step is the transport update in which finite-difference approximations to the integral advection equations are used to update the velocity, density and internal energy. 
Reconstruction at the cell boundaries is performed in a second order manner with a van Leer limiter \citep[see][]{vanleer77}. 

\subsection{Boundary Conditions}
At the stellar boundary we apply standard outflow boundary conditions \citep[see][]{stone92}. At the disk boundary we used two different approaches. For the steady simulations described in \S 4 the density and internal energy were held fixed, while for the simulations with explicit time variability described in \S 5 the density was varied according to a prescribed time-dependent function and the internal energy was varied such that the ghost cells maintained constant entropy.

\subsection{Simulation Parameters}
The simulations were performed in a dimensionless manner, however, we present the results here in dimensional form in order to compare to real systems. Since the flows are scale-free, the specific values of the density and temperature do not matter inside the simulation.

The material at the foot-point in the disk is assumed to have a density of $3 \times 10^{11}~ \mbox{cm}^{-3}$, this is taken from  the analytic accretion solutions of \citet{adams12} for a standard accretion rate of  $1\times10^{-8}$~M$_{\astrosun}$~yr$^{-1}$ and temperature of $10,000~\mbox{K}$ due to heating from UV radiation.
Because the cross-sectional area of the accretion column is a function of chosen multipole geometry, the `standardized $\dot{M}$' values reported in this work are mass fluxes at the star scaled to the equivalent mass flux through an area of one $\mbox{cm}^2$ at the surface of the disk. Mass accretion rates are scaled this way to facilitate comparisons between the two simulated geometries which would otherwise be drawing material from unequal sized magnetic footprints. Converting to the more observationally interesting quantities of the mass flux at the surface of the star can be done by dividing the reported scaled mass flux by the dimensionless cross section of the accretion column, $A$ (see Figure \ref{fig:geometry}). For the two geometries discussed in this paper, $\Gamma = 0$ and $\Gamma = 10$, the ratio of the cross-sectional area of the column at the stellar surface to the disk is $5.14 \times 10^{-4}$ and $4.18\times10^{-5}$ respectively. The global stellar accretion rate in terms of a hot spot filling factor, $f$, and this standardized mass flux can be written as
\begin{equation}
\dot{M}_{global} = f(4\pi R_*^2) \frac{\dot{M}}{A}
\end{equation}
For reference, a young accreting star with $f = 0.03$, $R = 1.5 R_{\odot}$ with a mass flux of $\dot{M} = 1.0\times10^{-7} \mbox{g\,cm}^{-2}$ under the pure dipole construction will have an accretion rate of $1.3\times 10^{-9} M_{\odot} \mbox{yr}^{-1}$ or $1.5 \times 10^{-8}M_{\odot} \mbox{yr}^{-1}$ under the dipole plus octupole configuration. Both accretion rates are plausible for T Tauri stars \citep[][]{ingleby13}.

We can determine the degree to which the flow is magnetically controlled by considering the magnetic plasma $\beta$ -- the ratio of the gas pressure to the magnetic pressure.  For a CTTS with a surface magnetic field strength $B_{\rm dip}$ = 3kG \citep[e.g.,][]{johns-krull07} and an inner disk edge of 10 $R_\ast$, the parameter $\beta\ll1$ several scale heights up in the disc where the accretion flow is launched. This finding validates our assumption that the magnetic field is unperturbed by the flow. In the midplane of disc, however, the density is larger and the magnetic plasma $\beta$ can approach unity. As result, MHD effects could be important in the midplane and hence in setting the location of the truncation radius.

The simulation domain is separated into to 1024 cells (resolution tests are discussed below). 
We fix the ghost cell used to set boundary conditions at the disk to a height of $0.05R_{*}$ above the mid-plane of the disk. 
Because gravitational potential and velocity gradients are larger near the star, a non-linear grid spacing is adopted to ensure the accuracy of the simulation. The grid spacing is scaled using a power law such that the resolution near the star is higher than near the disk. 

The simulation was initialized with a very small negative velocity. Density throughout the simulation is initially set to the density at the disk. The internal energy of the simulation, $e$, is set such that the entropy is constant throughout the simulation. Transient behavior from initialization dissipates over the length of several stellar sound crossing times and the choice of initial flow profile is not important. Time steps were chosen within the simulation by enforcing a maximum Courant number of 0.5. Additionally, the time-step was not allowed to increase by more than 30\% between subsequent steps.
 
\section{RESULTS: SEARCHING FOR SPONTANEOUS VARIABILITY}
As introduced above, previous analytic work suggested that a constraint can be placed on the index $n$ for steady-state solutions that pass smoothly through the sonic point and reach nearly free-fall speeds near the star \citep{adams12}.
If $n$ does not satisfy this constraint, the system may be unable to reach this type of solution. This effect could generate spontaneous variability within the accretion column, and could therby explain some of the short-term variability observed in TTS. This variability would be predicted to occur when the polytropic index $n$ is sufficiently small (corresponding to a stiff equation of state), i.e., $n < \ell + 3/2$, where $\ell$ is the highest order multipole with significant contributions to the magnetic field and in the inner limit where $r \rightarrow 0$. For the case where the stellar magnetic field is a pure dipole ($\ell = 1$, $\Gamma = 0$), the critical value for $n$ is $2.5$. 

\begin{figure}[h]
\centering
\includegraphics[width = .4\paperwidth]{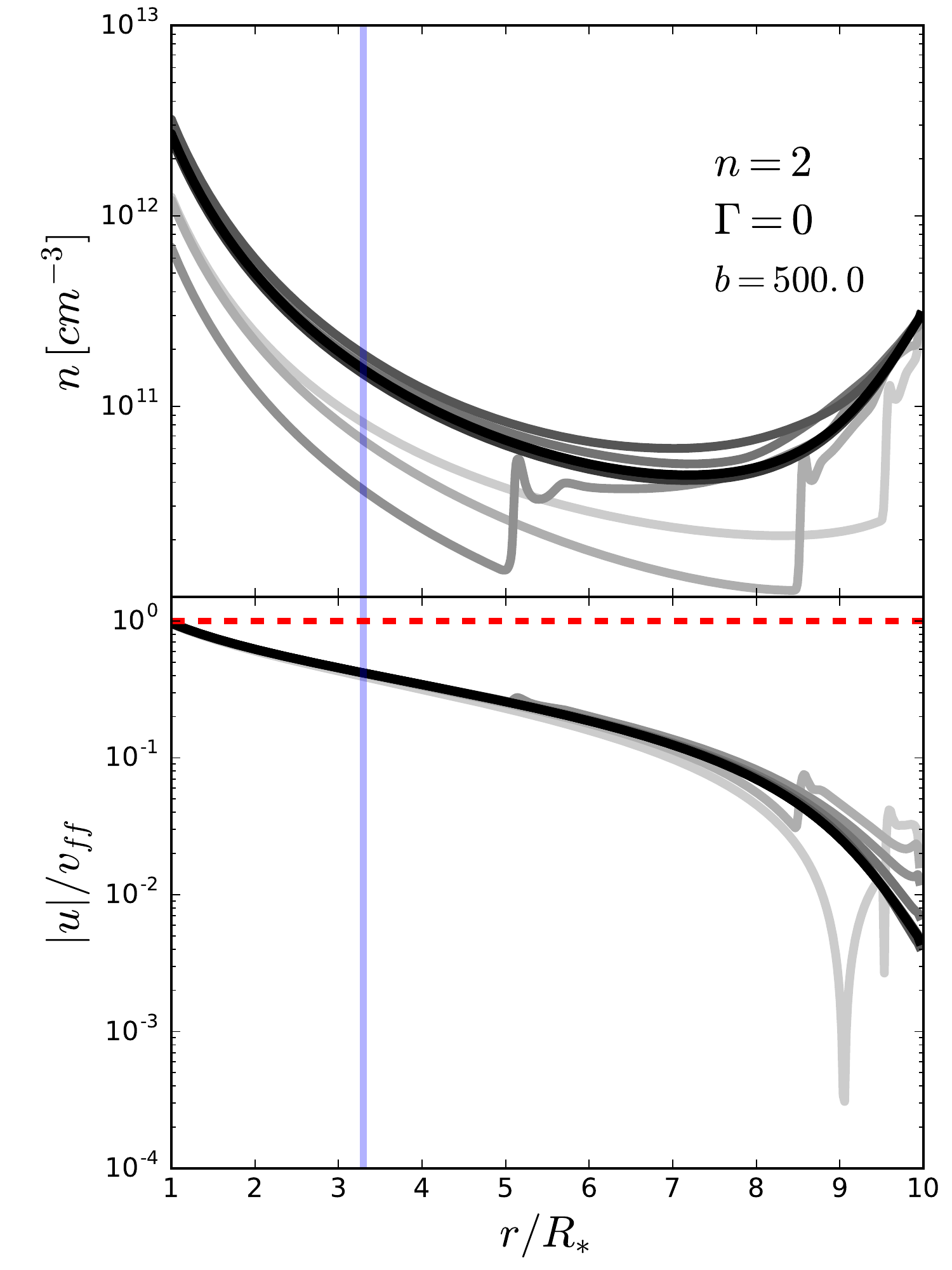}
\includegraphics[width = .4\paperwidth]{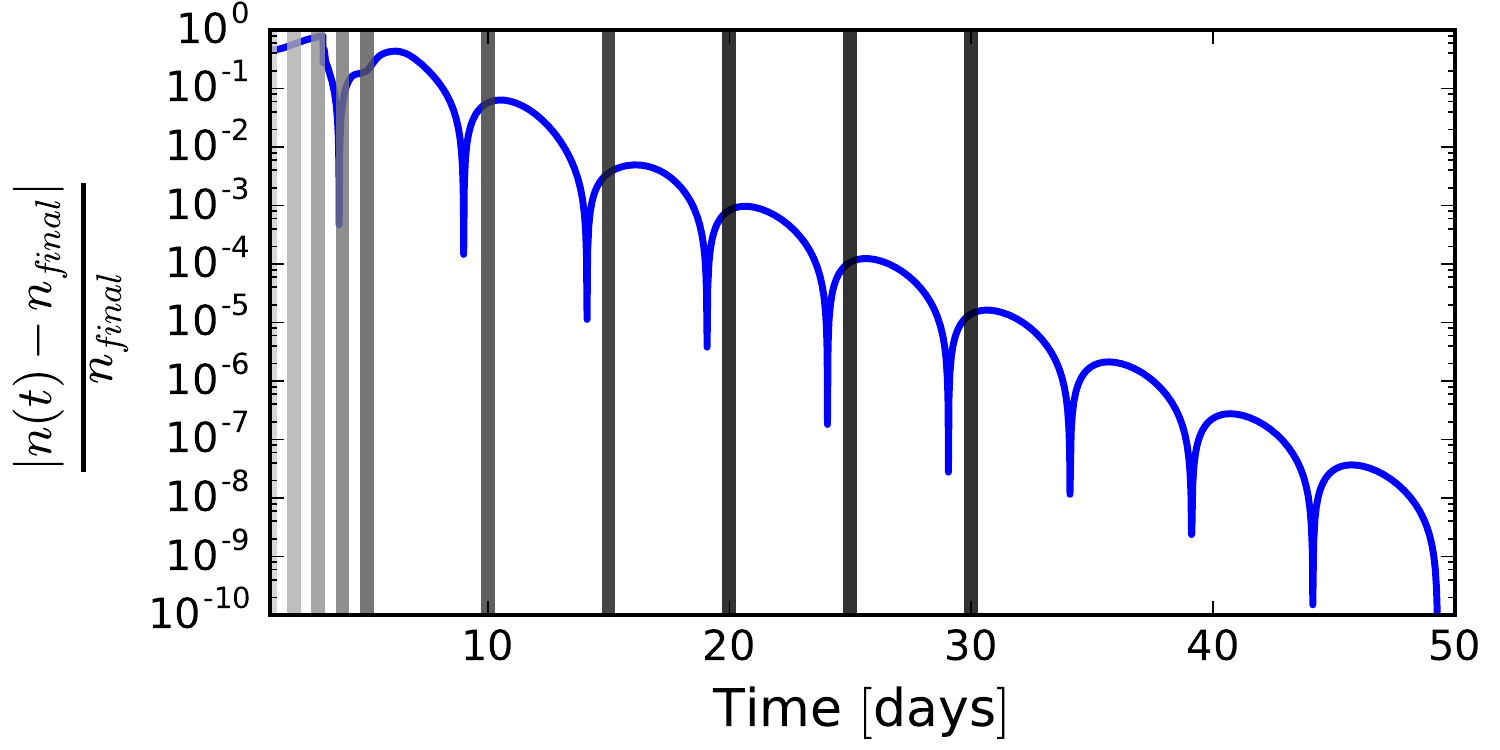}
\caption{\textit{Top panels:} Snapshots of density and speed for a transonic simulation under a dipole magnetic field configuration. The value of $n$ in this simulation is below the predicted critical value of $n = 2.5$, but the flow still converges to steady state. The snapshots are colored such that lines darken as time increases. The dashed red line shows the free-fall velocity. \textit{Bottom panel:} Absolute difference between the density as a function of time and the density in the final time step at the location of the blue line shown in the top panels. The vertical lines correspond to the snapshots shown in the top panel. The simulation oscillates and converges toward the steady state solution with an exponentially decaying amplitude.}
\label{fig:steady}
\end{figure}

To test this prediction, a set of simulations were performed above and below the critical value of the polytropic index with other parameters chosen to be representative of those for CTTS. The simulation set spanned $b = [200, 350, 500]$ and $n$ in the range 2.0--4.7 with a step size of 0.3 (although, lower $n$'s and $b$'s were explored -- see \S \ref{sec:subsonic}).
Transient effects from initialization dissipated over the course of a few sound crossing times ($\sim\,$days).
After these effects subsided, the simulations converged to a steady state and the flow was able to smoothly pass through the sonic point and reach free fall speeds for all combinations of $b$ and $n$ in this set. 
The condition of steady-state was confirmed by differencing the output velocity information between time steps, yielding accelerations in the regime of numerical rounding error. In order to be sure this is not a numerical issue, additional tests of a steady state flow are described in \S 4.1. Snapshots of an example simulation with values of $n = 2.0$ with $b = 500$ converging to a steady state solution can be seen in Figure \ref{fig:steady}. The absolute value of the residuals from differencing the density at a chosen point and the density at the same point in the final time step in the simulation is also shown in Figure \ref{fig:steady}. Velocity and density approach the final value with oscillatory behavior with a linearly decaying amplitude in semi-log space as time progresses, indicating the simulation is indeed converging towards a steady state solution. 

A second set of simulations was used to simulate systems with octupole + dipole stellar magnetic field geometry using the same distribution of values for $b$ and $n$. 
The addition of octupole components increases the amount of compression that occurs near the star, further heating the gas and raising pressure.
Spontaneous variability for this field line geometry is predicted to occur when $n < 4.5$.
The relative contribution of the octupole moment compared to the dipole moment for the magnetic field, $\Gamma$, was fixed to a value of 10 for this set of simulations.
These simulations were analyzed using the same method as the pure dipole case, and returned similar results; the simulations converged to a steady state after the initial transient behavior subsided. Again, the velocity was able to reach free fall speeds for each combination of $b$ and $n$.

\subsection{Subsonic Simulations}\label{sec:subsonic}
A subset of more extreme simulations with low masses and stiff equations of state were unable to reach a steady state solution because the flow failed to remain supersonic after the initial transient phase subsided.
The depth of the stellar potential well ($b$) determines where the sonic point is located along the column, and if it is small enough, the sonic point will lie within the star.
In this case, slow ($|u| < 0.1 v_{ff}$) time-variable solutions dominate the flow (analogous to the known time-variable ``breeze'' solutions \citealt{Velli94,DelZanna98,owen16}).
These solutions occur when $b$ is small, the compression of the flow is high (from octupole magnetic field components), and the flow is heated significantly ($n$ is small) and the sonic point criterion cannot be satisfied in the region between the inner edge of the disc and the stellar surface. 

The original criterion for non-free-fall flow has the form $n<\ell+3/2$ and thus does not depend on the depth $b$ of the gravitational potential well. On the other the hand, the above results show departures from free-fall flow with sufficiently small values of $b$. We can understand this behavior as follows. The incoming flow fails to reach free-fall speeds when the pressure term does not decrease fast enough relative to the gravitational term as the radius decreases. Under conditions where $n<\ell+3/2$, the pressure term will always dominate in the limit $\xi\to0$. As we have shown, however, the small dynamic range of the flow (the disc edge is only an order of magnitude larger than the stellar radius) does not generally allow for the flow to reach the regime (small $\xi$) where the pressure dominates, so the flow retains nearly free-fall speeds.  As shown in equations (24 -- 26) of \cite{adams12}, the depth $b$ of the gravitational potential well appears as a coefficient in the terms under comparison. As a result, when $b$ is large, the gravitational term dominates over a larger dynamic range (in $\xi$) and free-fall speeds can be reached even when $n<\ell+3/2$. For small values of $b$, the gravitational term is less effective at enforcing free-fall speeds and departures can arise (as discussed above). 

Nonetheless, these time-variable subsonic solutions are unlikely to present themselves in the region of parameter space realized by CTTS. Furthermore, observed signatures associated with shocks and high velocity flows at the surface of the star also indicate these solutions are not occurring. Although we reject these subsonic solutions as realistic models for CTTS, they may be relevant for systems with lower masses, e.g., planets and brown dwarfs, should they possess strong octupole (or higher order) magnetic field components. We caution this conjecture needs to be investigated in more detail before any predictions can be made.

$\,$
\bigskip

\subsection{Tests}

\begin{figure}[!h]
\centering
\includegraphics[width = .4\paperwidth]{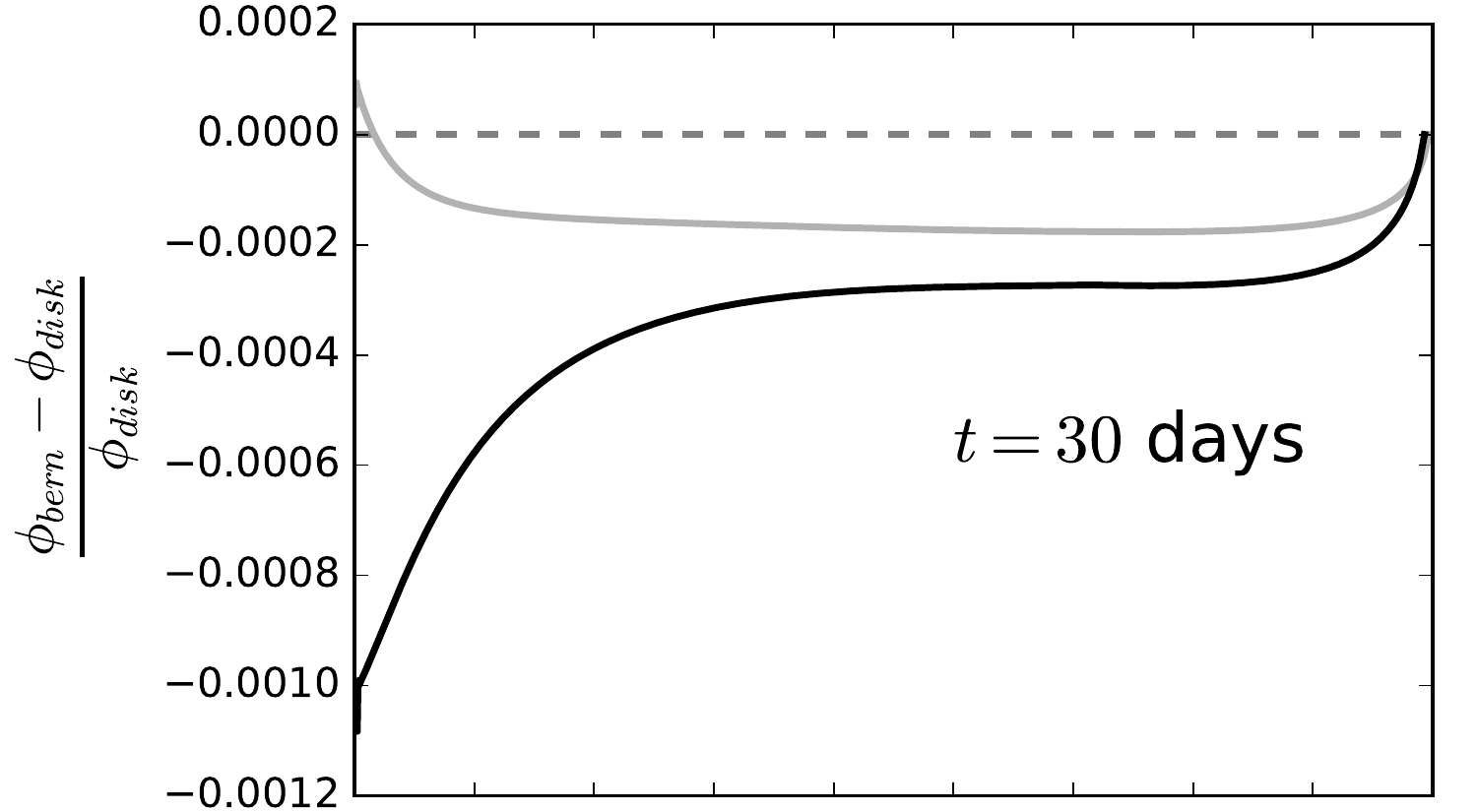}
\includegraphics[width = .4\paperwidth]{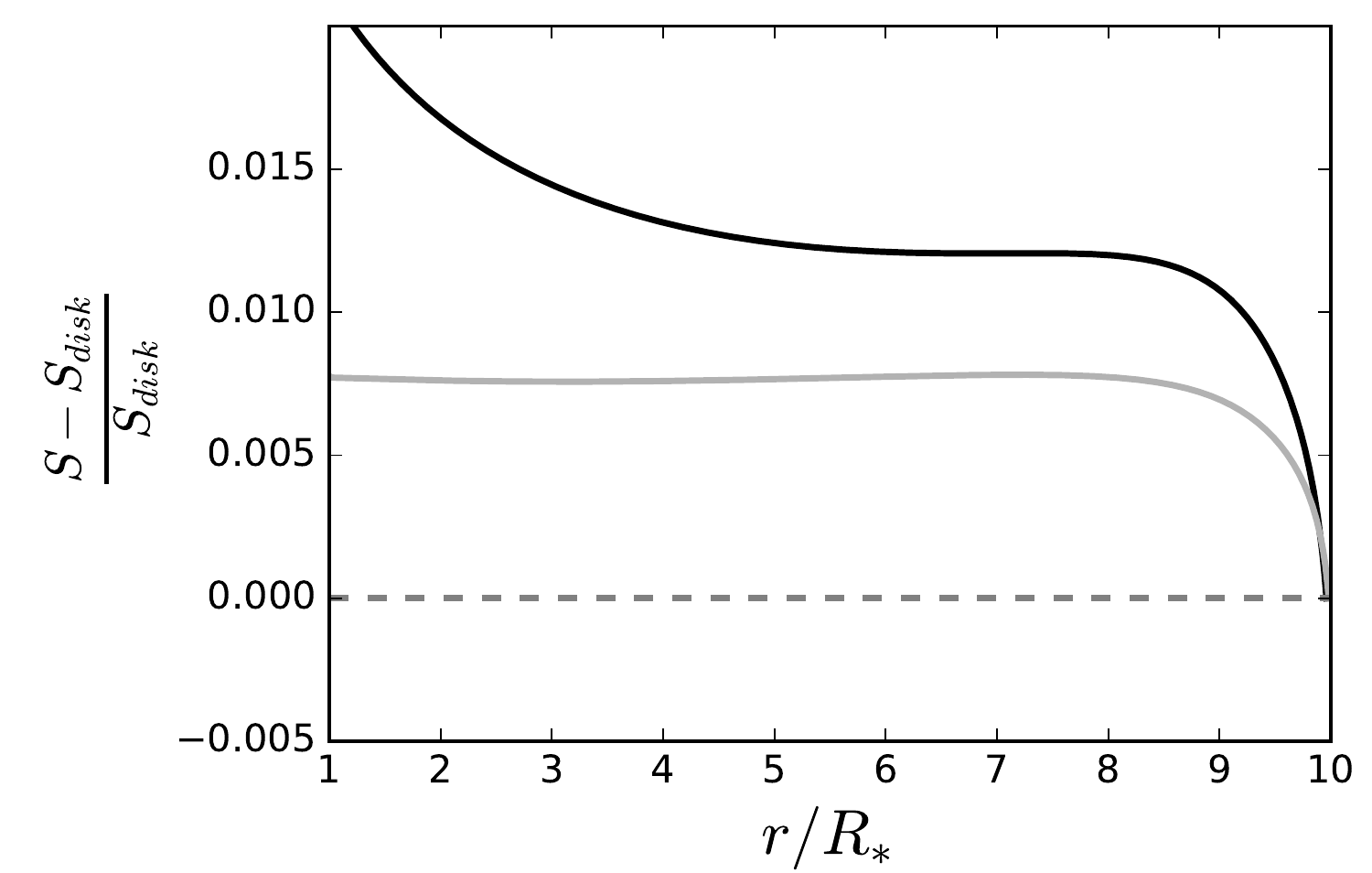}
\caption{\textit{Top panel}: Normalized relative difference in the Bernoulli potential at $t = 30$ for the simulation shown in Figure \ref{fig:steady} in solid black. A higher resolution simulation, shown in grey, was run with 2048 cells and shows smaller deviations from the expected constant potential. \textit{Bottom panel}: Normalized relative difference between entropy throughout the simulation and entropy at the disk.}
\label{fig:bern}
\end{figure}

Numerous tests were performed to check that the simulations were producing accurate results to ensure that the lack of spontaneous variability in the transonic solutions was not a spurious result from a numerical error.
In any steady-state flow the Bernoulli potential is constant; the Bernoulli potential of the flow was calculated and was found to converge to a nearly constant value after initial transient effects had dissipated. 
Figure \ref{fig:bern} shows the difference between the potential at the disk and the Bernoulli potential along the field line for the simulation shown in Figure \ref{fig:steady}.
Although there are deviations from a constant value along the field line, they are small and the magnitude of the deviations is in part dependant on the resolution of the simulation.
Under a resolution of 1024 cells, the maximum deviations are on the order of $0.1\%$.
The largest deviations from a constant potential generally occur near the star where the velocity of the flow is largest and the divergence of the magnetic field lines is steepest.
A second simulation is also shown in Figure \ref{fig:bern} in which the number of cells has been doubled to 2048, yielding maximum deviations from potential at the disk on the order of $0.02\%$.

The construction of our numerical scheme does not actively ensure that entropy is conserved to machine precision, as we explicitly work with internal energy. This provides another method to test the simulation, as the level of entropy in each cell should converge to a constant value once transient behavior has subsided under an adiabatic equation of state. 
The entropy of each cell was calculated and then normalized to the value of entropy at the inner edge of the disk.
Entropy remained roughly constant throughout the system with the largest deviations on the order of $1.5\%$. The bottom panel of Figure \ref{fig:bern} shows the normalized entropy throughout the simulation shown in Figure \ref{fig:steady}.
The same simulation run at the higher resolution is also shown in that panel with significantly smaller deviations, indicating that these deviations from constant entropy will continue to decrease with increasing resolution.

Finally, an analytic solution exists for the isothermal case. This provides a way to test the portions of the simulation not associated with the energy fluid equation. 
The results for a simulation using dipole geometry run under an isothermal equation of state is compared against the analytic solution in the Appendix. As shown in Figure \ref{fig:isothermal}, the simulation closely agrees with the analytic result. 

\subsection{Lack of Variability}

All of the transonic simulations run with a constant gas disk density converged to steady solutions with no indication of spontaneous variability, or lack of near free-fall velocities close to the stellar surface. 
The tests of the non-isothermal simulations in addition to the comparison of an isothermal flow against the analytic result suggest that the simulation is behaving physically. Instead, the assumptions that went into the prediction of a constraint on the possible values for $n$ by \citet{adams12} may be insufficient for realistic systems. 
In particular, those authors note that the assumption of working in the limit where $\xi \rightarrow 0$ may not hold where the dynamic range of the system is not sufficiently large. For example in CTTs the radial range is only about a factor of 10 from the stellar surface to the disc's edge.  
The constraint placed on the polytropic index is based on how the thermal pressure compares to the ram pressure in the innermost regions of the system. If the flow is to reach free-fall speeds, the ram pressure must dominate over the thermal pressure, which is the case for all of the supersonic simulations. In the cases where $n$ has a value lower than the predicted constraint, the dynamic range of $\xi$ does not appear to be large enough for the density term to dominate since the minimum value of $\xi$ is necessarily fixed at 1 (the stellar surface).

\section{RESULTS: DISK DRIVEN VARIABILITY}

Following the inability to explain short-term accretion variability through spontaneous variability in the accretion column in the transonic solutions, we ran sets of simulations using time variable boundary conditions. 
3D magnetohydrodynamic simulations have shown that density in the inner disk can be highly variable due to turbulence \citep{romanova12}.
As an approximation to these naturally occurring density perturbations, the density at the inner edge of the disk was modulated manually.
Two driving functions were used, sinusoidal oscillations and Gaussian pulses, both on top of a baseline density. These driving functions were chosen for convenience, allowing us to study a well posed problem, but are unlikely to represent the true density variations in real systems, which are likely to be stochastic.
The internal energy, and thus the pressure, at the boundary is set such that the cells at the boundary are varied adiabatically or isothermally in the isothermal models. The simulation is initialized in the same manner as before and is allowed to evolve for 15 days before the driving functions are initiated to avoid transients that occur upon initialization. It is shown in the bottom panel of Figure \ref{fig:steady} that relative departures from the steady state solution are on the order of $10^{-2}$ after this amount of time.

In these models, we choose the dimension-less parameters to correspond to a typical CTTS. These models were constructed with the following stellar properties: $M_{*} = 0.5\, M_{\astrosun}$, $R_{disk} = 10\, R_{*}$, and $R_{*} = 1.5\, R_{\astrosun}$. This gives us a value of $b\sim750$. Again, the temperature at the disk is assumed to be 10000K and at a density of $3\times 10^{11}\mbox{cm}^{-3}$.
The parameter space formed by $\Gamma$, $n$, the amplitude of the perturbation $A$, and the timescale of the perturbation was explored to examine the behavior of the accretion column under a variety of conditions. 

\subsection{Isothermal Sinusoidal Driving}

In the first set of simulations with driven accretion, the accretion column is assumed to be isothermal and the density at the inner edge of the disk was forced to oscillate sinusoidally with a fixed period $P$, and amplitude $A$, written as
\begin{equation}
\rho_{disk}(t) = \rho_1[A(\sin(\frac{2\pi}{P}t) +1) + 1].
\end{equation}
Simulations were run with P = [0.5, 1, 2, 4, 8] days, $\Gamma = [0,10]$ and $A=[1,2,3]$. 

\begin{figure}[!h]
\centering
\includegraphics[width = .4\paperwidth]{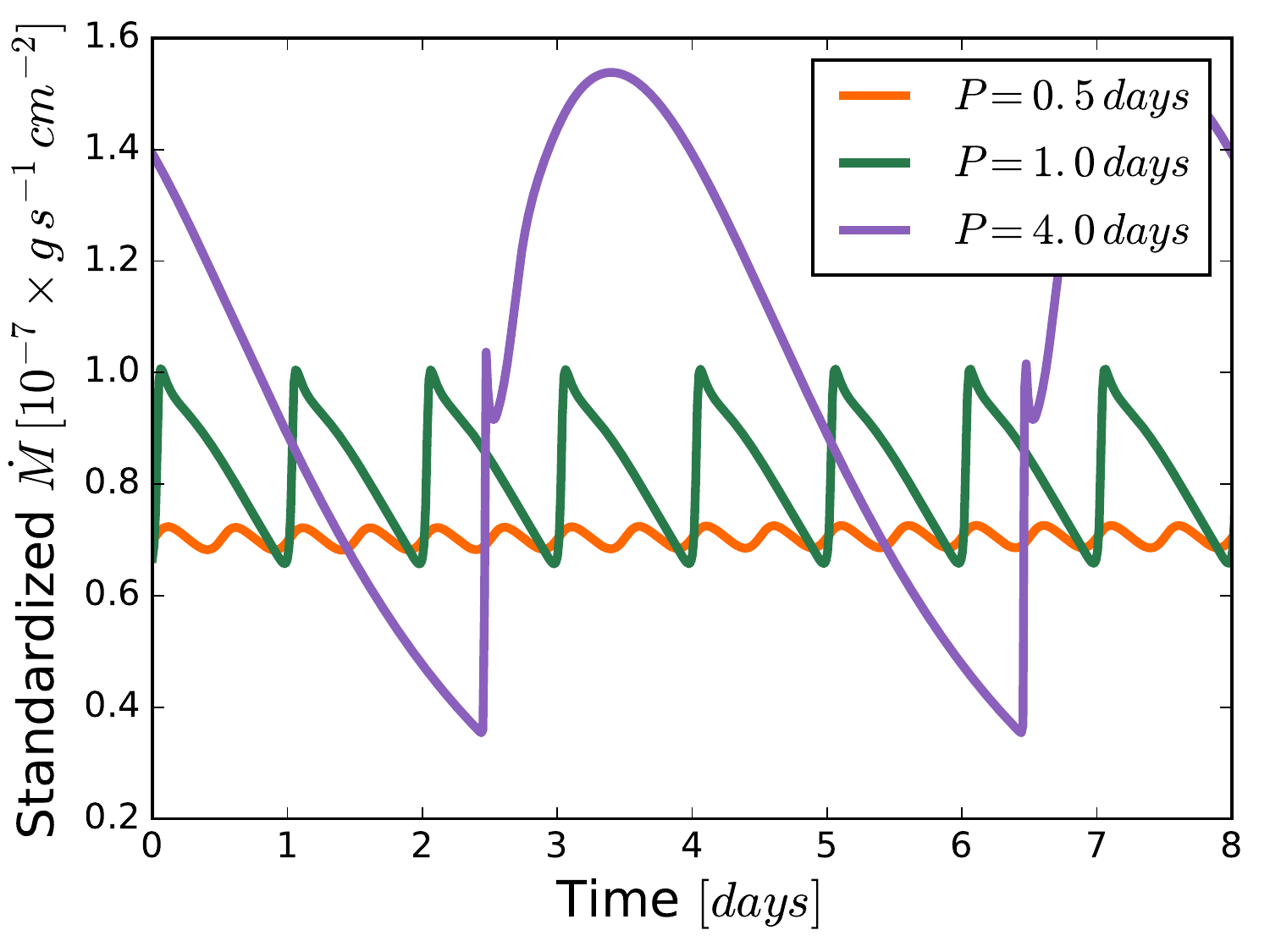}
\caption{Mass accretion rates near the surface of the star for isothermal simulations with different driving frequencies, written in terms of an equivalent mass flux at the disk. These simulations were run with oscillations with $A = 3$ under a pure dipole field configuration. Simulations with periods longer than roughly one sound crossing time form shocks in the outer regions of the accretion column that then propagate along the field line. Simulations with shorter periods do not shock, and show decreased accretion variability amplitudes compared to the longer period oscillations for the same value of $A$.}
\label{fig:isomdot}
\end{figure}

The behavior of the flow fell into three regimes depending on the driving period of the simulation.
Simulations of isothermal flows with driving periods shorter than about a day (roughly the stellar sound crossing time) showed smaller amplitudes than the driving function near the stellar surface. This is because the flow can not react fast enough to the changes in density in the outer boundary, i.e. the flow cannot adjust on a time-scale shorter than the sound crossing time. 
The over-densities in the accretion column are mostly time averaged out by the time that they collide with the star. The mass accretion rate of a simulation displaying this behavior with $P =  0.5$, $A = 3$, and $\Gamma = 0$. is shown by the orange line in Figure \ref{fig:isomdot}.

The second regime of behavior was observed in simulations with periods roughly one day or longer.
As the density is modulated at the inner edge of the disk, weak shocks form and then propagate along the field line. Figure \ref{fig:isomdot} shows two simulations in this regime in green and purple.
These shocks form because the velocity of the flow is slow in the outer regions of the accretion column, allowing material to build up during periods of high density in the disk, and then fall at once in the form of a higher density shock onto the star.  
The period of the accretion rate oscillations at the star matches the driving period of the boundary condition for all sets of simulation parameters. Finally, if the driving period becomes much longer than the flow time-scale (tens of days), the flow evolves steadily along the steady-state solutions.

\begin{figure}[!h]
\centering
\includegraphics[width = .4\paperwidth]{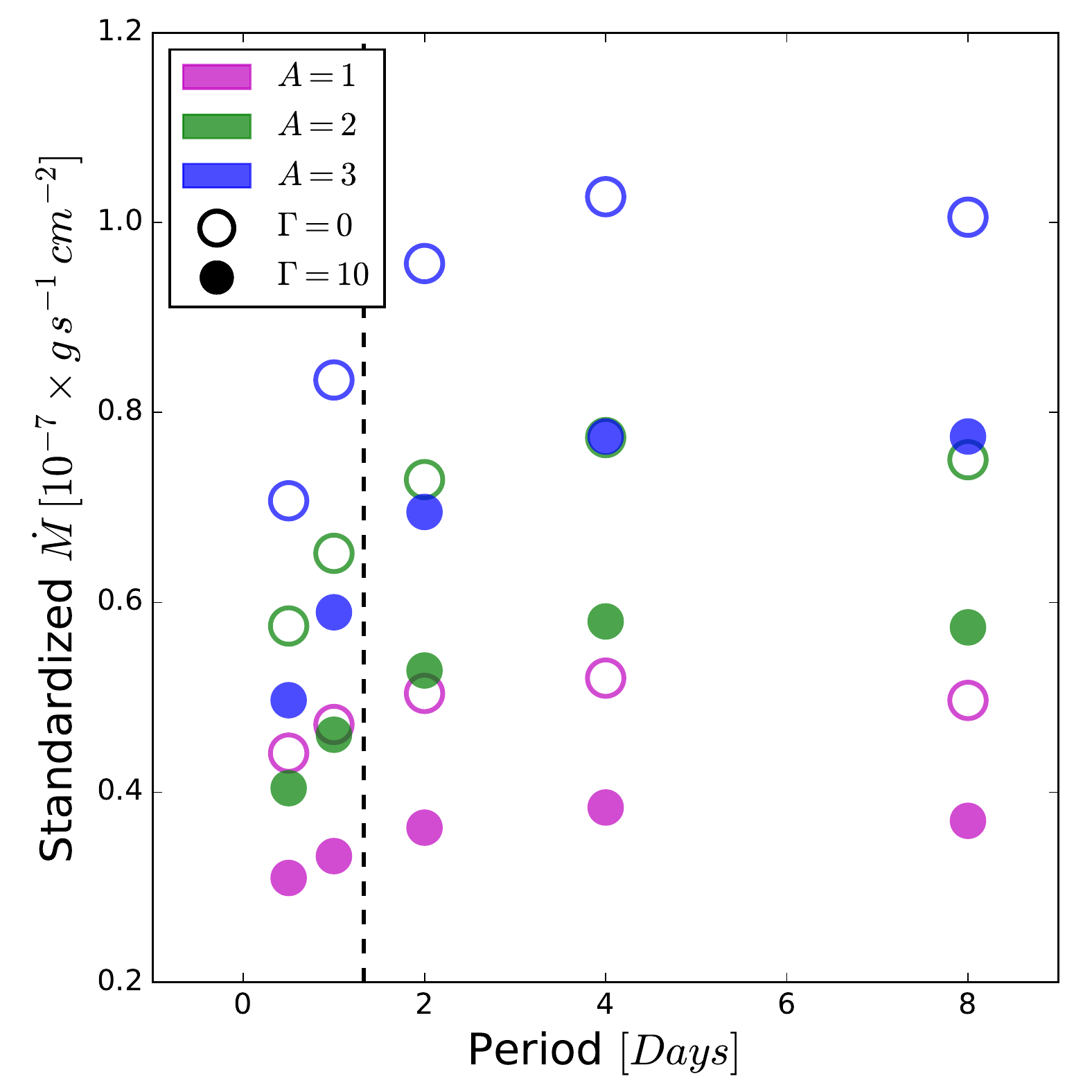}
\caption{Time averaged accretion rates written in terms of an equivalent mass flux at the disk, for isothermal simulations with sinusoidally driven boundary conditions. Different colors show driving amplitudes, $A$. Circles are simulations under a pure dipole field geometry while squares show simulations that have octupole components in addition to dipole components with $\Gamma = 10$.}
\label{fig:iso_avgmdot}
\end{figure}

In these driven accretion simulations, density is the quantity manually controlled at the outer boundary, which allows the mass loss rate to change as a function of model parameters. 
The averaged mass accretion rate at the star for all of the isothermal simulations is shown in Figure \ref{fig:iso_avgmdot}. Simulations with larger amplitudes show higher mass accretion rates, which is expected because the mean density in the outer regions of the accretion column is higher for larger values of $A$. Systems driven with shorter periods in most cases have lower accretion rates. 
This is in part due to sharp peaks in the density causing pressure gradients both inward and outward, that prevent matter from flowing onto the columns, and in some cases can even lead to flow away from the star. As the driving period increases, the pressure gradient become shallower.
Simulations run with magnetic fields comprised of dipole and octupole components show lower accretion rates than simulations run under the pure dipole field, in agreement with the analytic steady state solutions presented by \citet{adams12}.
In this construction of the magnetic field lines with $R_{disk} = 10R_{*}$, the net centripetal and gravitational force along the field line is away from the star in the outer regions of the accretion column for both the pure dipole and octupole + dipole configurations. This contributes to slow velocities in these regions.

\subsection{Non-isothermal Sinusoidal Driving}
\begin{figure}[!h]
\centering
\includegraphics[width = .4\paperwidth]{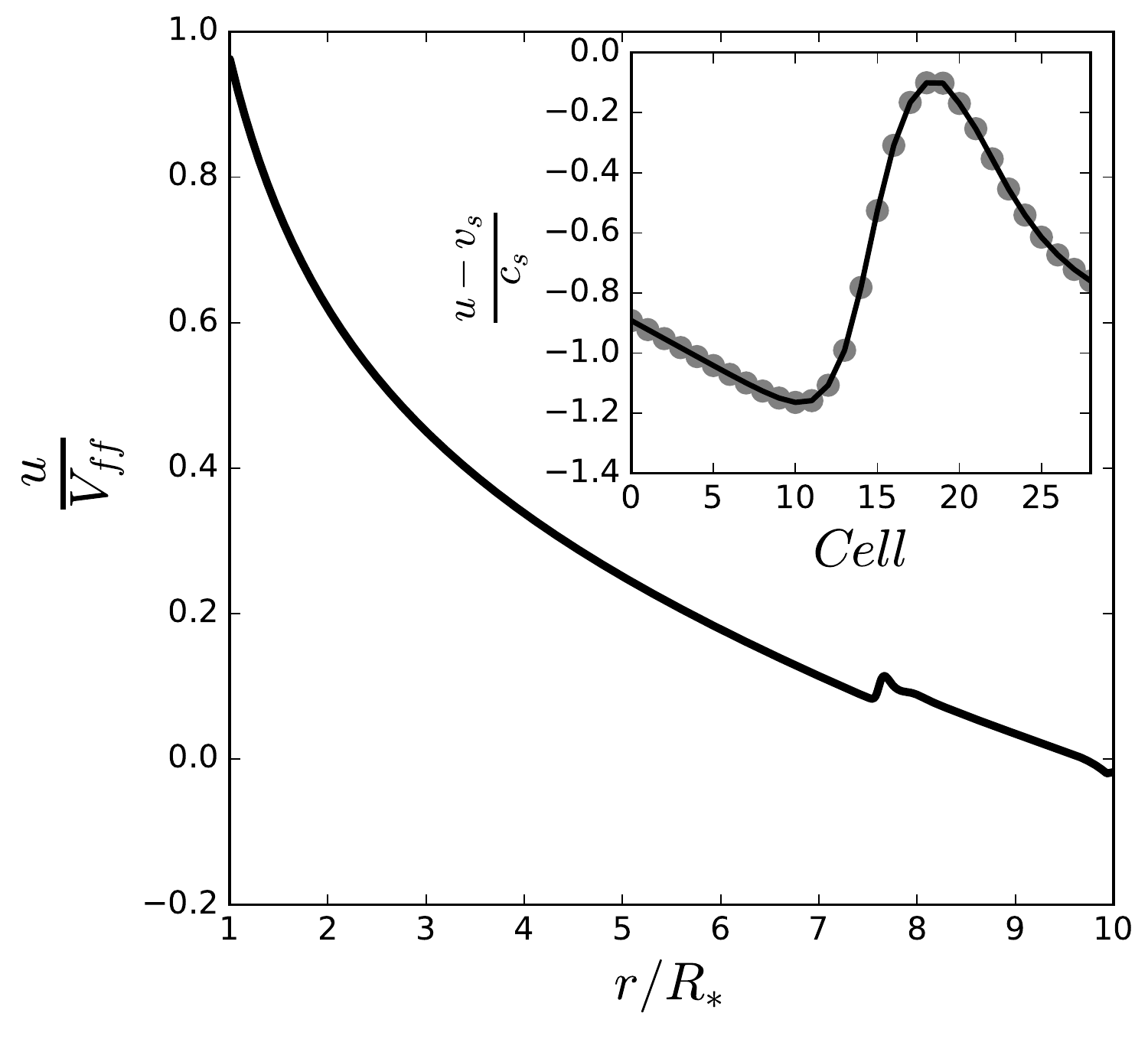}
\caption{Velocity snapshot in the accretion column for a simulation with sinusoidally driven outer boundaries. The simulation shown has values of $n = 3.5$, $P = 4.0$ days, $A = 3$, and $\Gamma = 0$. A shock forms in the outer regions of the simulation as material piles up, and then falls onto the star. The inset shows velocity in the reference frame of the shock in terms of the sound-speed. The grey dots in the inset show the location of the edge of each cell, which is where the velocity information is stored on the staggered mesh grid. Similar discontinuities also form in the isothermal simulations with driven accretion.}
\label{fig:discont}
\end{figure}

Another set of simulations with sinusoidal driving density perturbations at the inner edge of the disk were run with values of n = [3.5, 4.5, 5.5] using the same set of periods and amplitudes as the isothermal case. These simulations exhibit much of the same behavior as the isothermal simulations.
Similar to the isothermal case, weak shocks form in the outer regions of the accretion column in simulations with longer periods ($P > 1$ day).
Figure \ref{fig:discont} shows a snapshot of a simulation with a shock traveling along the accretion column. The shock speed for the inset plot was found using the Rankine–Hugoniot jump condition derived from the continuity equation using densities and velocities measured on either side of the shock.

\begin{figure}[!h]
\centering
\includegraphics[width = .4\paperwidth]{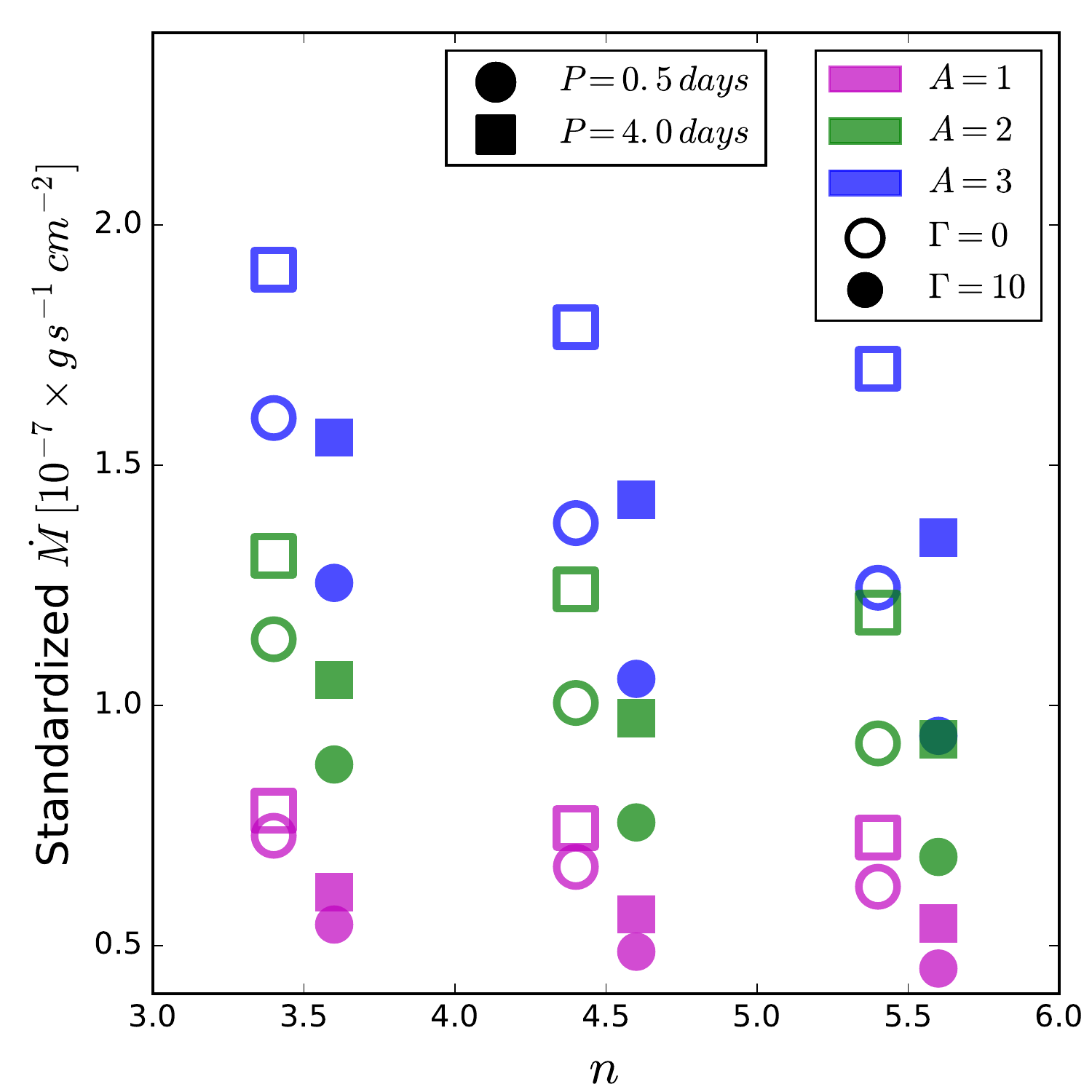}
\caption{Time averaged accretion rates for non-isothermal sinusoidally driven simulations as a function of $n$, $A$, $\Gamma$ and $P$. Open symbols show simulations with periods of 0.5d, and closed symbols show simulations with 4.0d periods. The points are shown slightly shifted from their true values of $n = [3.5, 4.5, 5.5]$ for clarity.}
\label{fig:noniso_avgmdot}
\end{figure}

The time averaged accretion rate for the non-isothermal simulations is shown in Figure \ref{fig:noniso_avgmdot}. Simulations with both short $0.5$d and longer $4.0$d oscillations are plotted to show both regimes of shocking and non-shocking behavior. For all simulations, the average accretion rate decreases as the accretion column becomes closer to isothermal.
Changing $A$, $\Gamma$ and $P$ reveals the same patterns as the isothermal models shown in Figure \ref{fig:iso_avgmdot}.

\subsection{Gaussian Density Pulses}
To isolate the effects that a single accretion burst would have on conditions near the stellar surface and test how long it takes the accretion column to relax from a perturbation, simulations were run with boundary conditions that manually vary the density at the disk-magnetosphere footprint as a Gaussian in time. This causes a density pulse to travel along the magnetic field line until it collides with the star, which for appropriate choices of the time-scale is a shock.
The density at the inner edge of the disk during the pulse is described as
\begin{equation}
\rho_{disk}(t) = \rho_1(1 + A e^{-\frac{t^2}{2\sigma^2}})
\end{equation}
where $A$ is the amplitude of the driving Gaussian and $\sigma$ represents the timescale of the pulse. 

\begin{figure}[!h]
\centering
\includegraphics[width = .4\paperwidth]{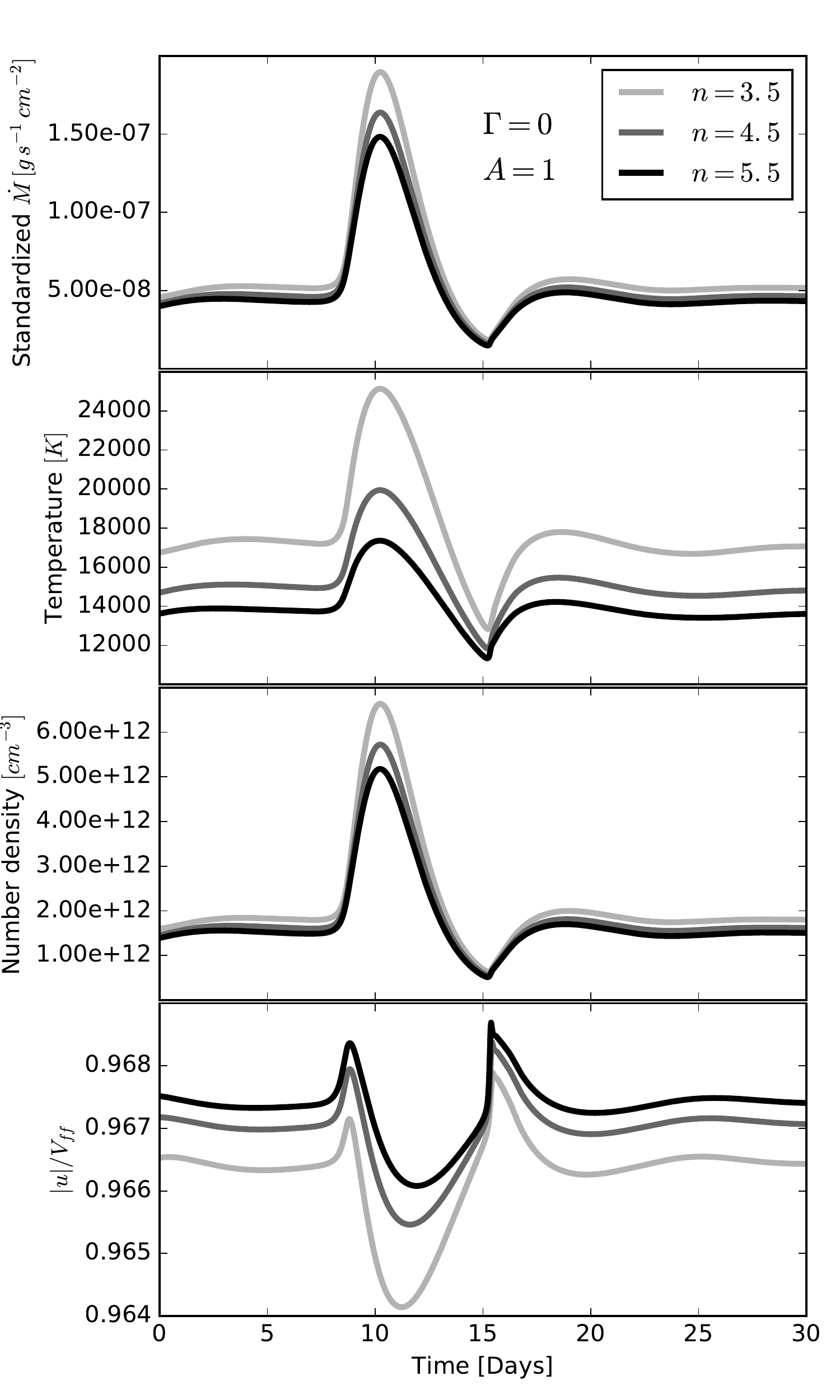}
\caption{Time dependant behavior for the calculations with a Gaussian density perturbation with varying values of the polytropic index. The top panel shows the accretion rate measured at the stellar surface. The next three panels show the pre-shock temperature, density and velocity measured at the stellar surface.}
\label{fig:vary_n}
\end{figure}

Figure \ref{fig:vary_n} shows the properties of the flow measured near the surface of the star as a function of time for several values of $n$, including the mass accretion rate, temperature, density, and velocity. The Gaussian density pulses shown have $\sigma = 1$ and $A = 3$. As discussed in section  5.1, the amount of material available for accretion onto the star is a function of the parameters describing the system, even if the time averaged density in the boundary remains the same.

\begin{figure}[!h]
\centering
\includegraphics[width = .4\paperwidth]{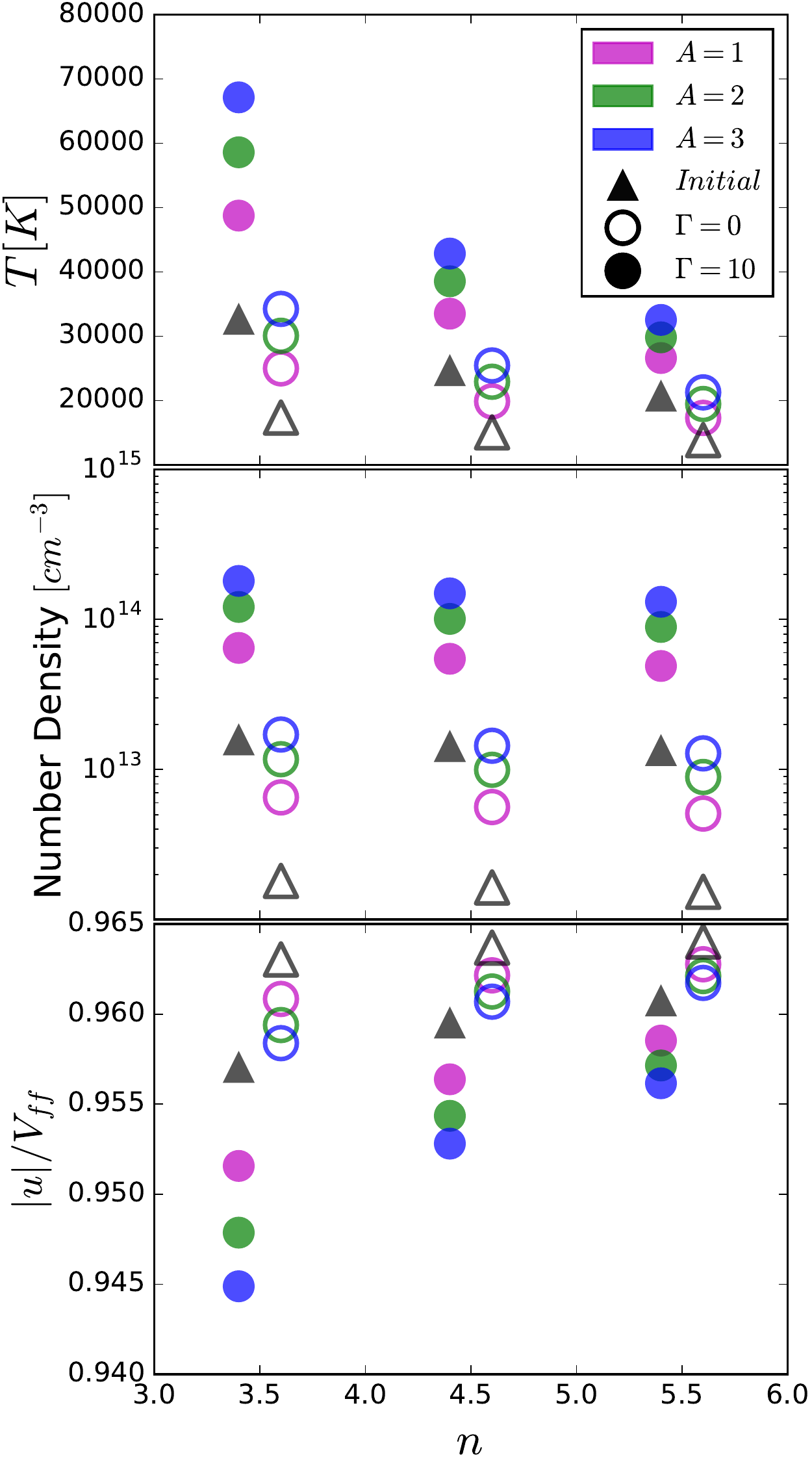}
\caption{Values of the pre-shock temperature, density and velocity taken near the stellar surface during the peak of the density pulse for the Gaussian driven boundary conditions. The points are shown slightly shifted from their values of $n = [3.5, 4.5, 5.5]$ for clarity.}
\label{fig:maxvals}
\end{figure}

The temperature, density, and velocity were measured near the star during the passage of the density pulse for a set of simulations with $A = [1,2,3]$, $\Gamma = [0,10]$, and $n = [3.5, 4.5, 5.5]$, the results of which are shown in Figure \ref{fig:maxvals}. The initial values of $T$, number density and $u$ for the steady state solutions, described in Section 4, are also shown in gray. As expected, the polytropic index is important in determining the temperature of the gas. As the polytropic index increases, the temperature of the gas decreases for all simulations and approaches the temperature of the upper layers of the gas disk (10000K). The density during and before the pulse is decreased in the simulations with higher polytropic indexes, but the effect is less dramatic than the effect on the temperature within the column.
The velocity near the star gradually increases with increasing values of $n$, but does not quite reach the free-fall velocity.

The geometry of the magnetic field lines also plays a critical role in the environment near the star. The field lines converge faster near the star for the octupole + dipole geometry, causing a higher level of compression at small radii compared to the pure dipole configuration.
For the most isothermal simulations shown here ($n = 5.5$) while the pulse is passing through, the temperature difference between the pure dipole and the octupole + dipole configurations  is on the order of 15000K. During steady state, the difference is roughly 8000K.
Density increases by roughly an order of magnitude while the pulse is passing through.
The changes in velocity are quite small between the two field configurations compared to the changes in the density and temperature as the thermal energy is relatively insignificant compared to the kinetic energy of the flow. The largest change in the velocity between the two configurations for any of the simulations during the passage of the pulse is only $\approx 1.2\%$ of the free-fall velocity. 

The amount of time it takes the column to relax to its initial state provides a measure of the timescale between discernible accretion events.
This timescale was found to be a function of the polytropic index and the geometry of the magnetic field lines.
Reminiscent of the bottom panel in Figure \ref{fig:steady}, the absolute value of the residuals between density in a cell near the stellar surface and the same cell at the final time step after the Gaussian pulse has passed can be enclosed within a power-law envelope. 
The envelope was found for each set of parameters within the sample by taking a finite temporal derivative of the density at a cell near the star and fitting the residuals with a power law at the times of the zero crossings.

\begin{figure}[!h]
\centering
\includegraphics[width = .4\paperwidth]{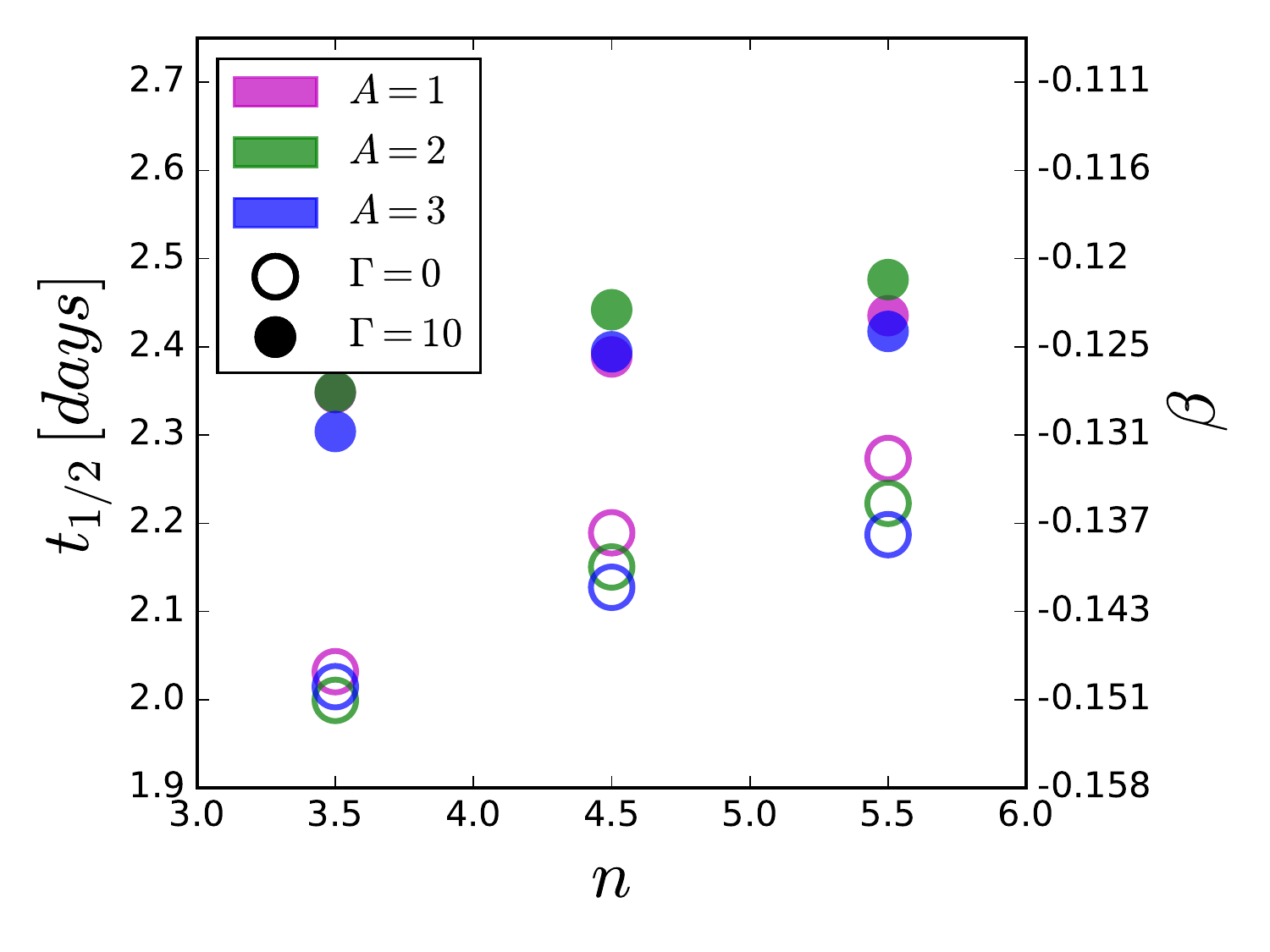}
\caption{Half life of the encompassing envelope of the oscillatory residuals from $|\rho(t) - \rho_{final}|$ near the surface of the star after a Gaussian pulse travels through the accretion column. The power-law index of the damped oscillation envelope is shown on the right axis. Gaussian pulses of different amplitudes ($A$) are shown in different colors, and the two cases of field line geometry are represented by open and closed circles.
The width of the driving Gaussian ($\sigma$) for these points is 2 days.} 
\label{fig:relax}
\end{figure}

The slope of the fit, $\beta$, and the half-life of the residuals is shown for all of the simulations in this sample in Figure \ref{fig:relax}.
The half-life for simulations under the octupole plus dipole magnetic field configuration is $\approx 10\%$ longer than those under a  pure dipole configuration. 
The polytropic index also plays a role in how quickly the accretion column relaxes. Simulations with smaller values of $n$ are more rapidly damped than more isothermal simulations. 
For comparison, two fully isothermal simulations with $A = 3$ were run for $\Gamma = 0$ and $\Gamma = 10$, and had $t_{1/2} = 2.3, 3.1$ days respectively. 
The amplitude of the initial Gaussian pulse plays a limited role in the relaxation timescale, which suggests a single slope/half-life may be sufficient for describing the behavior an accretion column undergoing events with similar timescales while the parameters governing the system remain constant.

\section{DISCUSSION}

We have discovered a new hydrodynamic mechanism that can lead to short-period, rapid changes in the accretion rate onto the star that may explain some of the short period accretion variability observed in CTTS. 
We find that smooth time-dependent variability in the inner disk that feeds the magnetospheric accretion column is amplified along the column. Under certain conditions these variations can shock in the accretion column, leading to rapid, large amplitude changes in the accretion rate onto the star. This is a promising mechanism to explain some of the variable behaviour observed in CTTS, as the inner disk that feeds the accretion column is likely to be variable. For example, density variations due to turbulence at the inertial scale have eddy turn-over times $\sim \Omega^{-1}$ \citep[e.g.,][]{Fromang2006}, which at the disk-magnetosphere boundary gives a time-scale of order a day with amplitudes of order unity. In further studies, we plan to model the exact observation signatures these shocks will produce; however, below we discuss our results in the context of recent observations and speculate about the connection between observations and our models.

A monitoring campaign of CTTS in NGC 2264 revealed variability in the light curves of most of the stars with significant UV excesses is dominated by short duration accretion bursts \citep{stauffer14}.
For all but one of these moderately accreting stars (which was systematically different than the others), the median rise and fall times were 0.4 and 0.5 days respectively, measured between the beginning or end of the burst and the center value. 
Qualitatively, this is comparable to the rise/fall behavior shown by some of the driven accretion scenarios presented here, e.g., the isothermal model driven using the sinusoidal boundary condition with a 1.0 day period shown in Figure \ref{fig:isomdot}. 
We also find shorter rise times in our simulations because of the shocking behavior presented in \S 5.1.
Although further investigation is required, we can speculate that the small asymmetry in the observed rise and fall times is because of the weak shocking behavior followed by a slower relaxation period. 
It is important to note that the boundary conditions used in these simulations are continuous, and the build up of material occurs within the active region of the simulation.
Most of the accretion burst dominated light curves presented by \citet{stauffer14} did not show periodic structure and instead bursted stochastically, providing inspiration for future simulations with density fluctuations at the outer boundary distributed randomly in time with varying amplitudes. In the case of the variability at the disk being turbulence driven, one naturally expects a range of time-scales rather than a simple periodic signal. 

In these simulations, the velocity approaches but never reaches the free-fall velocity measured from the inner edge of the disk 
\begin{equation}
    V_{ff} = \bigg(\frac{2GM}{R_{*}}\bigg)^{1/2} \bigg(1 - \frac{R_{*}}{R_{disk}}\bigg)^{1/2}
\end{equation}
because of the pressure support along the column.
This is relevant for popular methods of calculating mass accretion rates that assume the material reaches $V_{ff}$ as it strikes the surface of the star \citep[e.g.,][]{calvet98, gullbring98}.
Because the kinetic energy flux ($\rho u^3$) has such steep scaling with $u$, the difference between $V_{ff}$ and the velocities shown in Figure \ref{fig:maxvals} can result in calculated accretion rates that differ by up to $\sim 10\%$. 
Differences in velocities between simulations are all within $\sim 0.01 V_{ff}$ of each other, so the effect on the calculated accretion rate will be a roughly constant multiplicative factor.
As this is a nearly systematic effect for all models that assume $u \rightarrow V_{ff}$, this effect could easily be accounted for in future models of accretion shocks, although it does not drastically improve the accuracy of accretion rate measurements. Typical uncertainties in current measurements of accretion rates are generally much larger than 10$\%$. For example, uncertainties in $A_V$ alone can introduce a factor of 2 or larger \citep{ingleby15}. 

As shown in Figure \ref{fig:maxvals}, the addition of octupole components to the magnetic field drastically increases the density in the accretion column near the stellar surface.
The overall average accretion rate for this field configuration in our simulations is decreased slightly because of marginally slower velocities (see Figure \ref{fig:maxvals} and Figure \ref{fig:noniso_avgmdot}) and a smaller column cross section (Figure \ref{fig:geometry}). 
The increased density in accretion columns with dipole plus octupole fields leads to higher kinetic energy fluxes.
Higher kinetic energy fluxes lead to shock emissions that peak at shorter wavelengths \citep{calvet98}.
Combined with the smaller column cross section, this suggests that stars with significant octupole components are more likely to show evidence of hotter accretion spots with more limited surface coverage. 
Because the fractional changes between the initial and final velocity are small while pulses are passing through, changes in the energy flux at the surface of the star closely trace the density profile as a function of time. 
The thermal component of the energy flux within the column is essentially negligible for all the simulations shown here as the flow is highly super-sonic. 

\begin{figure}[!h]
\centering
\includegraphics[width = .4\paperwidth]{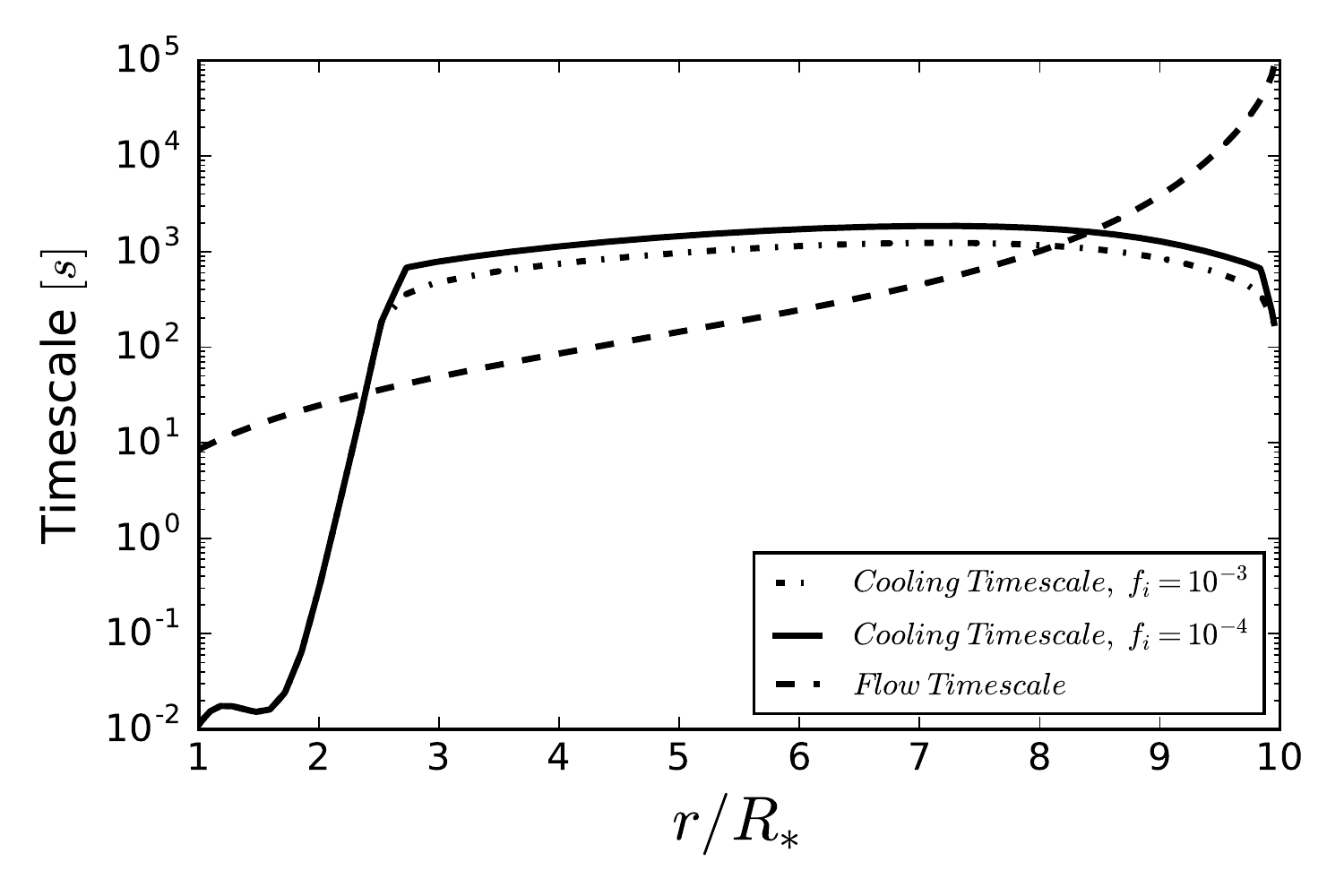}
\caption{Flow timescales and cooling timescales for the simulation shown in \ref{fig:steady} after transients have subsided. Two values of the ionization fraction are shown for regions in the simulation with temperatures under $10^4$K. The cooling timescale is faster or comparable to the flow timescale throughout the simulation, suggesting cooling is significant and should be included in future works. The cooling curves shown above are from \citet{schure09} and use solar metalicity values.}
\label{fig:cooling}
\end{figure}

Line emission associated with magnetospheric accretion columns has been observed \citep[e.g.,][]{johns95}. The simulation in its current form does not explicitly include radiative energy losses, which could play a role in determining the dynamics occurring within the accretion column, rather they are packaged up into in the polytropic index. As a first guess as to its importance, we can compare the optically thin cooling timescale to the flow-timescale. We note since this uses optically thin cooling it will provide a maximal estimate of the cooling rate, if any of the lines were to become optically thick, then the cooling rate would be reduced. Figure \ref{fig:cooling} shows the flow timescale, the time it takes for material to flow over a distance where the density changes significantly compared the cooling timescale derived from optically thin, solar metallicity, equilibrium cooling curves \citep{schure09}. The cooling is faster near the surface of the star where the magnetic field lines converge and material within the columns has been heated. Radiative cooling would likely drastically change the temperature structure seen in Figure \ref{fig:vary_n}. Figure \ref{fig:cooling} reveals cooling may also be significant near the magnetic footprint where the column connects with the disk where the flow velocities are slower. Cooling in the outer region of the accretion column would cause the flow to be more isothermal, affecting the formation process of the small shocks predicted in some of the driven accretion scenarios presented here. That said, discontinuities are present within the isothermal simulations presented here.

Because the cooling timescale is proportional to $\rho^{-1}$ for optically thin cooling and density in the simulation is normalized, the cooling timescale could be shifted up and down by changing the value for number density at the disk. 
The cooling curves shown here were generated assuming a density at the disk of $3\times10^{11} \mbox{cm}^{-3}$, yielding a surface mass flux of $1.26\times10^{-7}\mbox{g cm}^{-2}$.
The cooling timescale curve spans such a large amount of timescales that even if the density at the disk was decreased by a factor of 10, radiative cooling would still be important within the inner and outer regions of the simulation.
Because the mass accretion rate, and thus the density at the disk, can be highly variable, a range of values of the polytropic index are possible even for a fixed location along the accretion column. If the accretion rate is very low, e.g., $\dot{M} \approx 10^{-11} M_{\astrosun}/\mbox{yr}$, a more adiabatic equation of state is appropriate for the magnetic footprint at the disk ($n \rightarrow 1.5$ for a monatomic ideal gas). For a larger accretion rate, the gas is best approximated by an isothermal equation of state (n $\rightarrow \infty$). Using the simple $\rho^{-1}$ scaling for cooling timescales, estimates of when these regimes occur can be found using Figure \ref{fig:cooling} by scaling the cooling timescale curves inversely with density. The shape of the cooling curve is also itself a function of the polytropic index, so Figure~\ref{fig:cooling} only provides a rough guide.
Although the inclusion of radiative cooling is a logical next step in the development of this simulation, it would be a significant expansion on the current framework and thus will be left for future work. 

Finally, another source of heating or cooling that could arise is from MHD waves that are explicitly ignored in our treatment. Under an assumed magnetic field of 3kG the Alfv\'{e}n crossing time is significantly shorter than the flow crossing time ($\sim 20$ times shorter). This suggests that Alfv\'{e}n (and fast) waves\footnote{In the low $\beta$ flows we consider here the slow wave is essentially the same as sound waves.} are likely too fast to generate the sort of variability that creates these shocks; however, they may be able to transport energy and hence lead changes in the thermodynamics of the flow, which in principle could generate variability. Full MHD simulations are required to investigate the importance of MHD waves on the dynamics and thermodynamics.

The goal of this set of simulations was to expand 1D analytic steady state solutions into dynamic solutions allowing us to gain physical insight into their results. The simplicity of the 1D approach of these simulations also allows us to easily computed observable quantities.
However, the limitation of 1D comes at the cost of the loss of information associated with 2D and 3D simulations.
This includes the naturally occurring instabilities at the disk boundary \citep[e.g.,][]{romanova12}, which we have artificially replicated with driving functions.
We also lose information about anything that occurs in the toroidal direction. That said, the use of a 1D system allows for a higher degree of control over the system and allowed for the connections between theory and simulation. One possible future advancement that could be introduced within this 1D construction is the misalignment of the magnetic field and the stellar rotation axis.

\section{CONCLUSIONS}

We have performed 1D hydrodynamic simulations of accretion onto young stars under multipole magnetic field configurations aligned with the rotation axis, using both dipole and dipole plus octupole configurations. Static and time dependent inflow boundary conditions were used at the inner edge of the disk to investigate the possibility of spontaneous accretion variability and explore the effects of a time-variable accretion rate in the disk. Physical parameters that govern the flow were varied within the range expected in real accreting CTTS. The major conclusions of this work are summarized below. 

\begin{enumerate}

\item We found no spontaneous variability is generated within the column and the flow approaches steady state free-fall solutions for transonic solutions with values of $n$ lower than the constraint suggested by \citet{adams12}.
The analytic constraint on $n$ holds only in the limit of small radius $\xi\to0$. In contrast, we found it is likely that realistic young stellar systems lack the dynamic range in radius necessary for the constraint to come into effect.

\item By introducing time variability in the density at the inner edge of the disk, weak shocks can form in the outer regions of the accretion column and propagate towards the star resulting in rapid, high-amplitude changes in the accretion rate in systems with variability periods longer than roughly a day (of order the sound crossing time).
Variability driven with periods less than a day do not appear to shock, and the oscillations in density are averaged out, as the flow cannot respond fast enough to the changes.
With variability timescales much longer than the flow timescale, the flow profiles closely follow the steady solutions but slowly vary between them and no shocks form.  

\item The rapid increases in accretion rate seen in our simulations are qualitatively comparable to short duration accretion burst dominated variability observed in monitoring campaigns \citep{stauffer14}. Specifically, the shocks in the accretion columns naturally result in fast accretion rate rise times as the shock hits the star, and slower fall times as the flow relaxes to steady-state on the order of the sound crossing time (Figure \ref{fig:isomdot}).

\item Simulations were run featuring single time dependent Gaussian density pulses to reveal how the accretion column reacts during both quiescent and high accretion states. 
Decreasing the polytropic index leads to higher temperatures and higher mass accretion rates. 
Magnetic fields with dipole plus octupole components lead to much higher densities near the surface of the star with only slightly lower velocities compared to pure dipolar fields. 

\item We find the outer and inner regions of the accretion column are likely to be isothermal while the central portion may not be.
Optically thin cooling timescales of material in the column indicate the temperature profile is more complex than a single adiabatic or isothermal equation of state can provide (Figure \ref{fig:cooling}). 
Additional work to incorporate radiative cooling into this simulation is one possible next step to producing a more physically accurate representation of this system.
\end{enumerate}

We have found short term variability is not spontaneously generated within the accretion column, at least for our 1D examination of the system under realistic parameters for CTTS. 
Instead, variability on $\sim$day timescales is likely driven by instabilities and turbulence near the magnetic footprints of the flow, which can be modulated by shocks forming in the accretion columns. 
The limitations of this study lie in the assumptions of a polytropic treatment of pressure and the 1D treatment of the accretion column. 
Despite these limitations, these simulations provide a readily interpretable link between analytic solutions, more complex simulations and observations. In the future we expect an approach of the kind developed in this work can be used to directly compute time-dependent observables, such as the UV excess. 

\bigskip

\section{ACKNOWLEDGEMENTS}

We thank the anonymous referee who's comments helped improve the manuscript. CER and CCE acknowledge that this material is based upon work supported by the National Science Foundation under Grant No. AST-1455042.
JEO is supported by NASA through Hubble Fellowship grant HST-HF2-51346.001-A, awarded by the Space Telescope Science Institute (which is operated for NASA under contract NAS5-26555). FCA acknowledges support from the NASA Exoplanets Research Program NNX16AB47G, and from the University of Michigan.

\appendix
\section{NUMERICAL DETAILS}
\subsection{Finite-Difference Equations}
In this section we describe the finite difference scheme used to solve the system of hydrodynamic equations and discuss the associated scale factors and ancillary functions. The majority of the difference equations and their implementation were detailed in \citet{owen16}; here we detail the changes that have been made to include the magnetospheric accretion geometry and the energy equation.
Variables are subscribed with the location on the grid (i.e. $i-1$).
Time dependant variables ($u$, $\alpha$, $\epsilon$, and $P$) are superscribed with the time step (i.e $n+1$).
Time independent variables are superscribed with $a$ or $b$, depending on if the variable is face centered or zone centered respectively \citep[see also][]{stone92}.
Recall in this staggered scheme, $\alpha$ and $\epsilon$ (and thus $P$) are zone centered while $u$ is face centered \citep[e.g][]{stone92,owen16}.

During the source step, the velocity is updated due to forces from pressure, gravity, rotation and viscosity.
The velocity is first updated using the following finite-differenced version of the momentum equation:

\begin{equation}
u^{n+1}_i = u^n_{i} - \frac{\Delta t}{h^a_{p,i}} \frac{2}{(\alpha^n_{i}+\alpha^n_{i+1})}\bigg(\frac{P^n_{i} - P^n_{i+1}}{p^b_{i} - p^b_{i+1}}\bigg)
- \frac{\Delta t}{h^a_{p,i}} \bigg(\frac{b}{(\xi_i^a)^2}\bigg) \bigg({(\xi_i^a})^{-5}\frac{f^a_i\cos{\theta^a_i}}{H}\bigg)
+ \omega \Delta t \xi^a_i \sin{\theta^a_i} (\hat{x}\cdot \hat{p}^a_i).
\end{equation}
where $\omega$ is a dimensionless parameter that measures rotation, 
\begin{equation}
\omega = \big(\frac{\Omega R_{*}}{a_1}\big)^2.
\end{equation}
The component of $x$ along the field line, $\hat{x} \cdot \hat{p}$, can be written as a function of $\xi$, $\theta$, and the ancillary functions $f$, $g$ \& $H$ as
\begin{equation}
\hat{x} \cdot \hat{p} = \frac{\xi^{-5}\sin\theta\cos\theta(f+g)}{\sqrt{H}}
\end{equation}

Expressions for the ancillary functions and scale factors are shown below. The velocity and internal energy are then updated using an artificial viscosity tensor. More discussion on this tensor and the velocity update itself are included in Appendix A of \citet{owen16}. 
The source step is finished with the addition of the compressional heating update for the internal energy, performed using the explicit finite difference expression shown in \citep{stone92}.

Derivatives of the scale factors for both the face centered and zone centered grids were computed by finite differencing using the following expressions

\begin{align}
\begin{split}
\frac{\partial h^a_{x,i}}{\partial p} = \frac{h^b_{x,i} - h^b_{x,i-1}}{p^b_i - p^b_{i-1}}\\
\frac{\partial h^b_{x,i}}{\partial p} = \frac{h^a_{x,i} - h^a_{x,i-1}}{p^a_i - p^a_{i-1}}.
\end{split}
\end{align}
\citet{owen16} showed at sufficient resolution, the derivatives of the scale factors found by finite differencing are accurate enough for these geometric constructions for use in these simulations. This is useful since analytically obtaining these derivatives can be cumbersome.

\noindent The functional forms of the scale factors for $p$, $q$ and $\phi$ can be written as
\begin{equation}
h_p = \xi^{5} [ f^2\cos^2\theta + g^2\sin^2\theta]^{-1/2}
\end{equation}
\begin{equation}
h_q = \frac{\xi^{4}}{\sin\theta} [ f^2\cos^2\theta + g^2\sin^2\theta]^{-1/2}
\end{equation}
\begin{equation}
h_{\phi} = \xi \sin\theta
\end{equation}
\noindent where $f$ and $g$ are ancillary functions defined as
\begin{align}
\begin{split}
f \equiv \Gamma (5\cos^2\theta - 3) +2\xi^2\\
g \equiv \frac{3}{4}\Gamma (5\cos^2\theta -1) + \xi^2
\end{split}
\end{align}.

\noindent These ancillary functions recur often when using these coordinates and can be used in both the pure dipole case ($\Gamma = 0$) and the octupole plus dipole case. These functions can be combined to write $\abs{\nabla p}^2$, defined as $H$:

\begin{equation}
H \equiv \abs{\nabla p}^2 = \xi^{-10}[f^2\cos^2(\theta) + g^2\sin^2(\theta)].
\end{equation}

Finally, the transport step, which solves for changes in the time dependant variables due to fluid advection, follows the finite difference equations described in \citet{stone92}.

\subsection{Analytic vs. Simulation Comparison}

\noindent In order to be sure our numerical method is working as expected, we can compare our code to an analytic solution that exists for isothermal accretion.  A relatively simple solution for the velocity structure of the accretion column exists for the case of isothermal flow along dipole magnetic field geometry can be obtained using the procedure of \citep{cranmer04,adams12}. Following \citet{adams12}, an analytic solution for the velocity $u$ in the isothermal case can be written as 
\begin{equation}
u^2 = -W[-\lambda^{2}(q\xi)^{-6}(4-3q\xi)\exp(-2\varepsilon - \frac{2b}{\xi}-\omega q\xi^3)]
\end{equation}
where $W$ is the Lambert $W$ special function \cite[see also][]{cranmer04}, $\xi$ is the dimensionless radius, $b$ is the dimensionless depth of the potential well and $\varepsilon$ is an integration constant written as
\begin{equation}
\varepsilon = \frac{1}{2}\lambda^2-\frac{3}{2}bq.
\end{equation}
We note the Lambert $W$ function has two real branches, the `$0$' branch corresponds to sub-sonic flow and the `$-1$' branch to super-sonic flow. This solution for the velocity assumes the location where the field line intercepts the disk is the co-rotation radius such that $\omega = bq^3$. The integration constant $\lambda$ can be written in the transcendental form
\begin{equation}
\ln{\lambda} -\frac{\lambda^2}{2} = 3\ln{q\xi_s}-\frac{1}{2}\ln{(4-3q\xi_s)}-\frac{1}{2} +
b(-\frac{3}{2}q+\frac{1}{\xi_s}+\frac{1}{2}q^4\xi_s^3)
\end{equation}
where $\xi_s$ is the sonic point radii. Using the Lambert W function, the solution for $\lambda$ takes the form

\begin{equation}
\lambda^{2} = -W[-\exp{(-6\ln{q\xi_s}-\ln(4-3q\xi_s)
-1+b(3q+\frac{2}{\xi_{s}}+q^4\xi_s))}]
\end{equation}

where the sonic point $\xi_s$ is calculated using a numeric solution to the matching condition for the sonic point. The matching conditions is written as 
\begin{equation}
3\xi_s(\frac{8-5q\xi_s}{4-3q\xi_s})+3\omega q\xi_s^4 = 2b.
\end{equation}

An isothermal dipole simulation with $b = 250$, a disk radius of $\xi_{\mbox{d}} = 10$ and 1024 cells was run using a constant density outer boundary condition and outflow inner boundary conditions. The output from this simulation was then compared against the analytic solution shown above. For this value of $b$, the sonic point is at a radius of 8.71 stellar radii. A comparison of the analytic solution and the simulation result can be seen in Figure \ref{fig:isothermal}.

\renewcommand{\thefigure}{A\arabic{figure}}

\setcounter{figure}{0}

\begin{figure}
\label{fig:isothermal}
\epsscale{0.6}
\plotone{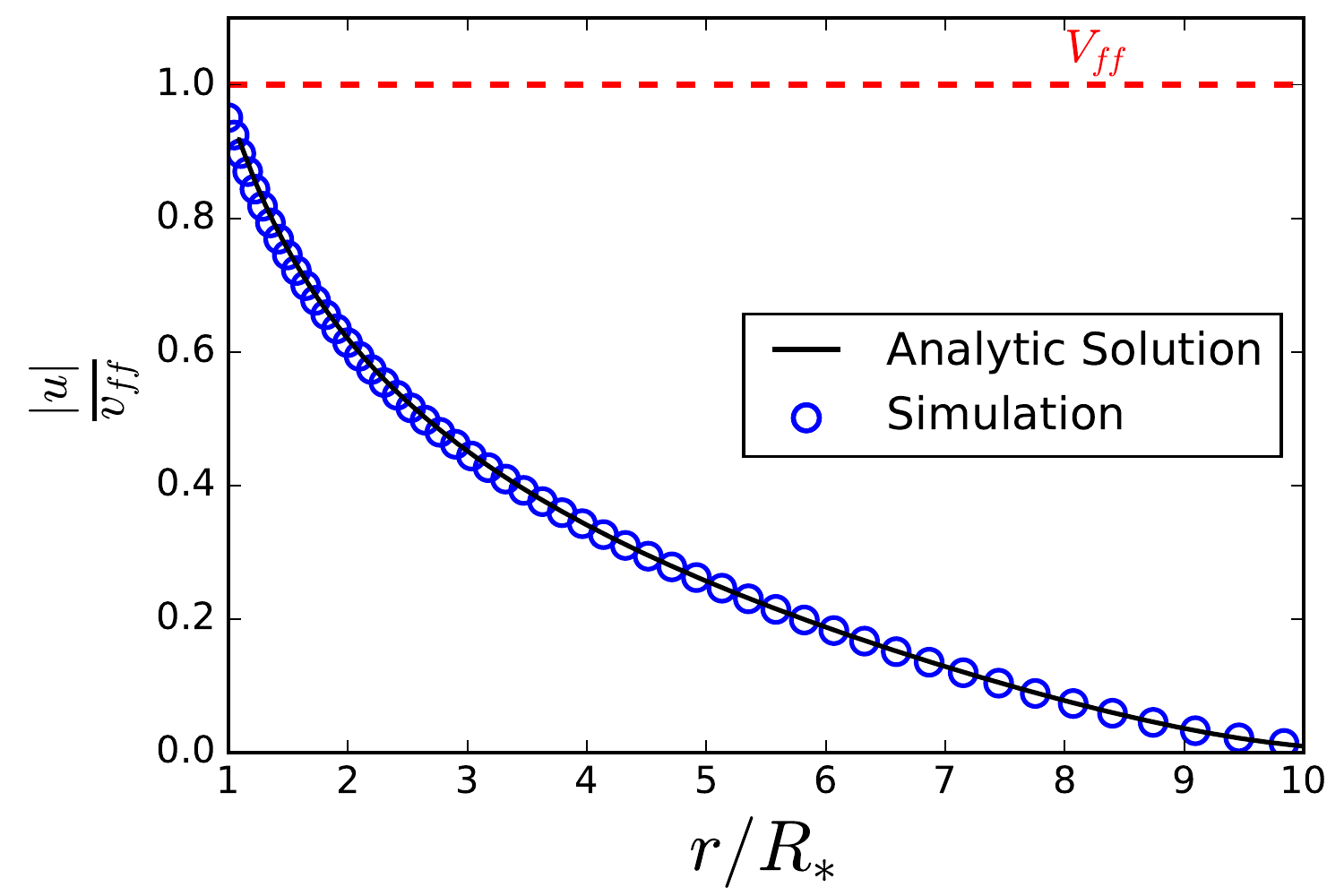}
\caption{Analytic solution to for the case of isothermal transonic accretion compared to the result from the simulation. The simulation result converges to this analytic solution, ensuring the simulation is returning physical results. Median deviations from the analytic solution are on the order of $0.7\%$. The true resolution of the simulation is 20 times finer than indicated by the number of open circles.}
\end{figure}

\bibliographystyle{apalike}
\bibliography{biblio}

\begin{thebibliography}{}

\bibitem[{Adams} and {Gregory}, 2012]{adams12}
{Adams}, F.~C. and {Gregory}, S.~G. (2012).
\newblock {Magnetically Controlled Accretion Flows onto Young Stellar Objects}.
\newblock {\em \apj}, 744:55.

\bibitem[{Alencar} et~al., 2010]{alencar10}
{Alencar}, S.~H.~P., {Teixeira}, P.~S., {Guimar{\~a}es}, M.~M., {McGinnis},
  P.~T., {Gameiro}, J.~F., {Bouvier}, J., {Aigrain}, S., {Flaccomio}, E., and
  {Favata}, F. (2010).
\newblock {Accretion dynamics and disk evolution in NGC 2264: a study based on
  CoRoT photometric observations}.
\newblock {\em \aap}, 519:A88.

\bibitem[{Andrews} et~al., 2009]{andrews09}
{Andrews}, S.~M., {Wilner}, D.~J., {Hughes}, A.~M., {Qi}, C., and {Dullemond},
  C.~P. (2009).
\newblock {Protoplanetary Disk Structures in Ophiuchus}.
\newblock {\em \apj}, 700:1502--1523.

\bibitem[{Ardila} et~al., 2013]{ardila13}
{Ardila}, D.~R., {Herczeg}, G.~J., {Gregory}, S.~G., {Ingleby}, L., {France},
  K., {Brown}, A., {Edwards}, S., {Johns-Krull}, C., {Linsky}, J.~L., {Yang},
  H., {Valenti}, J.~A., {Abgrall}, H., {Alexander}, R.~D., {Bergin}, E.,
  {Bethell}, T., {Brown}, J.~M., {Calvet}, N., {Espaillat}, C., {Hillenbrand},
  L.~A., {Hussain}, G., {Roueff}, E., {Schindhelm}, E.~R., and {Walter}, F.~M.
  (2013).
\newblock {Hot Gas Lines in T Tauri Stars}.
\newblock {\em \apjs}, 207:1.

\bibitem[{Bouvier} et~al., 1999]{bouvier99}
{Bouvier}, J., {Chelli}, A., {Allain}, S., {Carrasco}, L., {Costero}, R.,
  {Cruz-Gonzalez}, I., {Dougados}, C., {Fern{\'a}ndez}, M., {Mart{\'{\i}}n},
  E.~L., {M{\'e}nard}, F., {Mennessier}, C., {Mujica}, R., {Recillas}, E.,
  {Salas}, L., {Schmidt}, G., and {Wichmann}, R. (1999).
\newblock {Magnetospheric accretion onto the T Tauri star AA Tauri. I.
  Constraints from multisite spectrophotometric monitoring}.
\newblock {\em \aap}, 349:619--635.

\bibitem[{Calvet} and {Gullbring}, 1998]{calvet98}
{Calvet}, N. and {Gullbring}, E. (1998).
\newblock {The Structure and Emission of the Accretion Shock in T Tauri Stars}.
\newblock {\em \apj}, 509:802--818.

\bibitem[{Calvet} and {Hartmann}, 1992]{calvet92}
{Calvet}, N. and {Hartmann}, L. (1992).
\newblock {Balmer line profiles for infalling T Tauri envelopes}.
\newblock {\em \apj}, 386:239--247.

\bibitem[{Cohen} et~al., 2004]{cohen04}
{Cohen}, R.~E., {Herbst}, W., and {Williams}, E.~C. (2004).
\newblock {A Multiyear Photometric Study of IC 348}.
\newblock {\em \aj}, 127:1602--1621.

\bibitem[{Cranmer}, 2004]{cranmer04}
{Cranmer}, S.~R. (2004).
\newblock {New views of the solar wind with the Lambert W function}.
\newblock {\em American Journal of Physics}, 72:1397--1403.

\bibitem[{Del Zanna} et~al., 1998]{DelZanna98}
{Del Zanna}, L., {Velli}, M., and {Londrillo}, P. (1998).
\newblock {Dynamical response of a stellar atmosphere to pressure
  perturbations: numerical simulations}.
\newblock {\em \aap}, 330:L13--L16.

\bibitem[{Donati} et~al., 2011]{donati11b}
{Donati}, J.-F., {Gregory}, S.~G., {Alencar}, S.~H.~P., {Bouvier}, J.,
  {Hussain}, G., {Skelly}, M., {Dougados}, C., {Jardine}, M.~M., {M{\'e}nard},
  F., {Romanova}, M.~M., and {Unruh}, Y.~C. (2011).
\newblock {The large-scale magnetic field and poleward mass accretion of the
  classical T Tauri star TW Hya}.
\newblock {\em \mnras}, 417:472--487.

\bibitem[{Espaillat} et~al., 2011]{espaillat11}
{Espaillat}, C., {Furlan}, E., {D'Alessio}, P., {Sargent}, B., {Nagel}, E.,
  {Calvet}, N., {Watson}, D.~M., and {Muzerolle}, J. (2011).
\newblock {A Spitzer IRS Study of Infrared Variability in Transitional and
  Pre-transitional Disks Around T Tauri Stars}.
\newblock {\em \apj}, 728:49.

\bibitem[{Fromang} and {Papaloizou}, 2006]{Fromang2006}
{Fromang}, S. and {Papaloizou}, J. (2006).
\newblock {Dust settling in local simulations of turbulent protoplanetary
  disks}.
\newblock {\em \aap}, 452:751--762.

\bibitem[{Gregory} and {Donati}, 2011]{gregory11}
{Gregory}, S.~G. and {Donati}, J.-F. (2011).
\newblock {Analytic and numerical models of the 3D multipolar magnetospheres of
  pre-main sequence stars}.
\newblock {\em Astronomische Nachrichten}, 332:1027.

\bibitem[{Gregory} et~al., 2016]{gregory16}
{Gregory}, S.~G., {Donati}, J.-F., and {Hussain}, G.~A.~J. (2016).
\newblock {The multipolar magnetic fields of accreting pre-main-sequence stars:
  B at the inner disk, B along the accretion flow, and B at the accretion
  shock}.
\newblock {\em ArXiv e-prints}.

\bibitem[{Gullbring} et~al., 1998]{gullbring98}
{Gullbring}, E., {Hartmann}, L., {Brice{\~n}o}, C., and {Calvet}, N. (1998).
\newblock {Disk Accretion Rates for T Tauri Stars}.
\newblock {\em \apj}, 492:323--341.

\bibitem[{Hartmann}, 1998]{hartmannbook98}
{Hartmann}, L. (1998).
\newblock {\em {Accretion Processes in Star Formation}}.

\bibitem[{Hartmann} et~al., 1994]{hartmann94}
{Hartmann}, L., {Hewett}, R., and {Calvet}, N. (1994).
\newblock {Magnetospheric accretion models for T Tauri stars. 1: Balmer line
  profiles without rotation}.
\newblock {\em \apj}, 426:669--687.

\bibitem[{Ingleby} et~al., 2011]{ingleby11}
{Ingleby}, L., {Calvet}, N., {Bergin}, E., {Herczeg}, G., {Brown}, A.,
  {Alexander}, R., {Edwards}, S., {Espaillat}, C., {France}, K., {Gregory},
  S.~G., {Hillenbrand}, L., {Roueff}, E., {Valenti}, J., {Walter}, F.,
  {Johns-Krull}, C., {Brown}, J., {Linsky}, J., {McClure}, M., {Ardila}, D.,
  {Abgrall}, H., {Bethell}, T., {Hussain}, G., and {Yang}, H. (2011).
\newblock {Near-ultraviolet Excess in Slowly Accreting T Tauri Stars: Limits
  Imposed by Chromospheric Emission}.
\newblock {\em \apj}, 743:105.

\bibitem[{Ingleby} et~al., 2013]{ingleby13}
{Ingleby}, L., {Calvet}, N., {Herczeg}, G., {Blaty}, A., {Walter}, F.,
  {Ardila}, D., {Alexander}, R., {Edwards}, S., {Espaillat}, C., {Gregory},
  S.~G., {Hillenbrand}, L., and {Brown}, A. (2013).
\newblock {Accretion Rates for T Tauri Stars Using Nearly Simultaneous
  Ultraviolet and Optical Spectra}.
\newblock {\em \apj}, 767:112.

\bibitem[{Ingleby} et~al., 2015]{ingleby15}
{Ingleby}, L., {Espaillat}, C., {Calvet}, N., {Sitko}, M., {Russell}, R., and
  {Champney}, E. (2015).
\newblock {Using FUV to IR Variability to Probe the Star-Disk Connection in the
  Transitional Disk of GM Aur}.
\newblock {\em \apj}, 805:149.

\bibitem[{Isella} et~al., 2009]{isella09}
{Isella}, A., {Carpenter}, J.~M., and {Sargent}, A.~I. (2009).
\newblock {Structure and Evolution of Pre-main-sequence Circumstellar Disks}.
\newblock {\em \apj}, 701:260--282.

\bibitem[{Johns} and {Basri}, 1995]{johns95}
{Johns}, C.~M. and {Basri}, G. (1995).
\newblock {Hamilton Echelle Spectra of Young Stars. II. Time Series Analysis of
  H(alpha) Variations}.
\newblock {\em \aj}, 109:2800.

\bibitem[{Johns-Krull}, 2007]{johns-krull07}
{Johns-Krull}, C.~M. (2007).
\newblock {The Magnetic Fields of Classical T Tauri Stars}.
\newblock {\em \apj}, 664:975--985.

\bibitem[{Koenigl}, 1991]{koenigl91}
{Koenigl}, A. (1991).
\newblock {Disk accretion onto magnetic T Tauri stars}.
\newblock {\em \apjl}, 370:L39--L43.

\bibitem[{Koldoba} et~al., 2002]{koldoba02}
{Koldoba}, A.~V., {Lovelace}, R.~V.~E., {Ustyugova}, G.~V., and {Romanova},
  M.~M. (2002).
\newblock {Funnel Flows from Disks to Magnetized Stars}.
\newblock {\em \aj}, 123:2019--2026.

\bibitem[{Kulkarni} and {Romanova}, 2008]{kulkarni08}
{Kulkarni}, A.~K. and {Romanova}, M.~M. (2008).
\newblock {Accretion to magnetized stars through the Rayleigh-Taylor
  instability: global 3D simulations}.
\newblock {\em \mnras}, 386:673--687.

\bibitem[{Li}, 1996]{li96}
{Li}, J. (1996).
\newblock {Magnetic Interaction between Classic T Tauri Stars and Their
  Associated Disks}.
\newblock {\em \apj}, 456:696.

\bibitem[{Livio} and {Pringle}, 1992]{livio92}
{Livio}, M. and {Pringle}, J.~E. (1992).
\newblock {Dwarf nova outbursts - The ultraviolet delay and the effect of a
  weakly magnetized white dwarf}.
\newblock {\em \mnras}, 259:23P--26P.

\bibitem[{Mendoza V.}, 1966]{mendoza66}
{Mendoza V.}, E.~E. (1966).
\newblock {Infrared Photometry of T Tauri Stars and Related Objects}.
\newblock {\em \apj}, 143:1010.

\bibitem[{Muzerolle} et~al., 2001]{muzerolle01}
{Muzerolle}, J., {Calvet}, N., and {Hartmann}, L. (2001).
\newblock {Emission-Line Diagnostics of T Tauri Magnetospheric Accretion. II.
  Improved Model Tests and Insights into Accretion Physics}.
\newblock {\em \apj}, 550:944--961.

\bibitem[{Muzerolle} et~al., 2003]{muzerolle03}
{Muzerolle}, J., {Calvet}, N., {Hartmann}, L., and {D'Alessio}, P. (2003).
\newblock {Unveiling the Inner Disk Structure of T Tauri Stars}.
\newblock {\em \apjl}, 597:L149--L152.

\bibitem[{Muzerolle} et~al., 2009]{muzerolle09}
{Muzerolle}, J., {Flaherty}, K., {Balog}, Z., {Furlan}, E., {Smith}, P.~S.,
  {Allen}, L., {Calvet}, N., {D'Alessio}, P., {Megeath}, S.~T., {Muench}, A.,
  {Rieke}, G.~H., and {Sherry}, W.~H. (2009).
\newblock {Evidence for Dynamical Changes in a Transitional Protoplanetary Disk
  with Mid-Infrared Variability}.
\newblock {\em \apjl}, 704:L15--L19.

\bibitem[{Muzerolle} et~al., 1998]{muzerolle98}
{Muzerolle}, J., {Hartmann}, L., and {Calvet}, N. (1998).
\newblock {A Brgamma Probe of Disk Accretion in T Tauri Stars and Embedded
  Young Stellar Objects}.
\newblock {\em \aj}, 116:2965--2974.

\bibitem[{Natta} et~al., 2001]{natta01}
{Natta}, A., {Prusti}, T., {Neri}, R., {Wooden}, D., {Grinin}, V.~P., and
  {Mannings}, V. (2001).
\newblock {A reconsideration of disk properties in Herbig Ae stars}.
\newblock {\em \aap}, 371:186--197.

\bibitem[{O'dell} and {Wen}, 1994]{odell94}
{O'dell}, C.~R. and {Wen}, Z. (1994).
\newblock {Postrefurbishment mission Hubble Space Telescope images of the core
  of the Orion Nebula: Proplyds, Herbig-Haro objects, and measurements of a
  circumstellar disk}.
\newblock {\em \apj}, 436:194--202.

\bibitem[{Owen} and {Adams}, 2016]{owen16}
{Owen}, J.~E. and {Adams}, F.~C. (2016).
\newblock {Hot Jupiter breezes: time-dependent outflows from extrasolar
  planets}.
\newblock {\em \mnras}, 456:3053--3067.

\bibitem[{Romanova} et~al., 2012]{romanova12}
{Romanova}, M.~M., {Ustyugova}, G.~V., {Koldoba}, A.~V., and {Lovelace},
  R.~V.~E. (2012).
\newblock {MRI-driven accretion on to magnetized stars: global 3D MHD
  simulations of magnetospheric and boundary layer regimes}.
\newblock {\em \mnras}, 421:63--77.

\bibitem[{Schure} et~al., 2009]{schure09}
{Schure}, K.~M., {Kosenko}, D., {Kaastra}, J.~S., {Keppens}, R., and {Vink}, J.
  (2009).
\newblock {A new radiative cooling curve based on an up-to-date plasma emission
  code}.
\newblock {\em \aap}, 508:751--757.

\bibitem[{Shu} et~al., 1987]{shu87}
{Shu}, F.~H., {Adams}, F.~C., and {Lizano}, S. (1987).
\newblock {Star formation in molecular clouds - Observation and theory}.
\newblock {\em \araa}, 25:23--81.

\bibitem[{Shu} et~al., 1994]{shu94}
{Shu}, F.~H., {Najita}, J., {Ruden}, S.~P., and {Lizano}, S. (1994).
\newblock {Magnetocentrifugally driven flows from young stars and disks. 2:
  Formulation of the dynamical problem}.
\newblock {\em \apj}, 429:797--807.

\bibitem[{Stauffer} et~al., 2014]{stauffer14}
{Stauffer}, J., {Cody}, A.~M., {Baglin}, A., {Alencar}, S., {Rebull}, L.,
  {Hillenbrand}, L.~A., {Venuti}, L., {Turner}, N.~J., {Carpenter}, J.,
  {Plavchan}, P., {Findeisen}, K., {Carey}, S., {Terebey}, S.,
  {Morales-Calder{\'o}n}, M., {Bouvier}, J., {Micela}, G., {Flaccomio}, E.,
  {Song}, I., {Gutermuth}, R., {Hartmann}, L., {Calvet}, N., {Whitney}, B.,
  {Barrado}, D., {Vrba}, F.~J., {Covey}, K., {Herbst}, W., {Furesz}, G.,
  {Aigrain}, S., and {Favata}, F. (2014).
\newblock {CSI 2264: Characterizing Accretion-burst Dominated Light Curves for
  Young Stars in NGC 2264}.
\newblock {\em \aj}, 147:83.

\bibitem[{Stone} and {Norman}, 1992]{stone92}
{Stone}, J.~M. and {Norman}, M.~L. (1992).
\newblock {ZEUS-2D: A radiation magnetohydrodynamics code for astrophysical
  flows in two space dimensions. I - The hydrodynamic algorithms and tests.}
\newblock {\em \apjs}, 80:753--790.

\bibitem[{Ustyugova} et~al., 2006]{ustyugova06}
{Ustyugova}, G.~V., {Koldoba}, A.~V., {Romanova}, M.~M., and {Lovelace},
  R.~V.~E. (2006).
\newblock {``Propeller'' Regime of Disk Accretion to Rapidly Rotating Stars}.
\newblock {\em \apj}, 646:304--318.

\bibitem[{Valenti} and {Johns-Krull}, 2004]{valenti04}
{Valenti}, J.~A. and {Johns-Krull}, C.~M. (2004).
\newblock {Observations of Magnetic Fields on T Tauri Stars}.
\newblock {\em \apss}, 292:619--629.

\bibitem[{van Leer}, 1977]{vanleer77}
{van Leer}, B. (1977).
\newblock {Towards the Ultimate Conservative Difference Scheme. IV. A New
  Approach to Numerical Convection}.
\newblock {\em Journal of Computational Physics}, 23:276.

\bibitem[{Velli}, 1994]{Velli94}
{Velli}, M. (1994).
\newblock {From supersonic winds to accretion: Comments on the stability of
  stellar winds and related flows}.
\newblock {\em \apjl}, 432:L55--L58.

\end{thebibliography}

\end{document}